\documentclass[acmsmall,natbib=false,nonacm,authorversion]{acmart}

\AtBeginDocument{%
  \providecommand\BibTeX{{%
    Bib\TeX}}}

\setcopyright{acmcopyright}
\copyrightyear{2018}
\acmYear{2018}
\acmDOI{XXXXXXX.XXXXXXX}

\acmPrice{15.00}
\acmISBN{978-1-4503-XXXX-X/18/06}

\usepackage{algorithmic}
\usepackage{graphicx}
\usepackage{textcomp}
\usepackage{xcolor}
\usepackage{etoolbox}
\usepackage{multirow}
\usepackage{url}
\usepackage{multirow,booktabs}
\usepackage{subcaption}
\usepackage{wrapfig}
\usepackage{float}
\usepackage{float}
\usepackage{tcolorbox}
\usepackage[flushleft]{threeparttable}
\usepackage{relsize}
\usepackage[inline]{enumitem}
\usepackage{amsmath}

\definecolor{mygreen}{rgb}{0.0, 0.5, 0.0}

\newcommand{\revise}[1]{{\color{black}{#1}}}

\usepackage{tikz}

\AtBeginDocument{%
  \providecommand\BibTeX{{%
    \normalfont B\kern-0.5em{\scshape i\kern-0.25em b}\kern-0.8em\TeX}}}

\ccsdesc[500]{Computing methodologies~Artificial intelligence}
\ccsdesc[500]{Security and privacy}

\keywords{Statement-level vulnerability detection, Mutual information, and Contrastive learning.}

\newcommand{\smallsection}[1]{\noindent {\bf #1}.\hspace{1mm}}

\usepackage{soul}

\newcommand{\ourapp}{\textsc{LEAP}}

\newcommand{\rqone}{Can our \ourapp~approach identify vulnerable code statements in vulnerable functions in an unsupervised setting (where there is no ground truth of vulnerable code statements required in the training process)?}
\newcommand{\rqtwo}{Can our \ourapp~approach identify vulnerable code statements in vulnerable functions more accurately in a semi-supervised setting than in an unsupervised setting?}
\newcommand{\rqthree}{What are the contributions of the components of our \ourapp~approach (mutual information and our proposed cluster spatial contrastive learning)?}

\usepackage{amsmath,amsfonts,bm}

\def\eqref#1{equation~\ref{#1}}

\def\1{\bm{1}}

\def\vp{{\bm{p}}}

\DeclareMathAlphabet{\mathsfit}{\encodingdefault}{\sfdefault}{m}{sl}
\SetMathAlphabet{\mathsfit}{bold}{\encodingdefault}{\sfdefault}{bx}{n}

\begin{document}

\title{Statement-Level Vulnerability Detection: Learning Vulnerability Patterns Through Information Theory and Contrastive Learning}
\renewcommand{\shorttitle}{Statement-Level Vulnerability Detection Through Information Theory and Contrastive Learning}

\author{Van Nguyen}
\affiliation{%
  \institution{Monash University}
  \city{Clayton}
  \country{Australia}}
\email{van.nguyen1@monash.edu}
\authornote{Corresponding author: Van Nguyen. This research was done by Van Nguyen during his tenure as a Postdoctoral Research Fellow in the Department of Data Science and AI at Monash University, Australia. It was supported by the Defence Science and Technology Group’s Next Generation Technologies Program, Australia.}

\author{Trung Le}
\affiliation{%
  \institution{Monash University}
  \city{Clayton}
  \country{Australia}}
\email{trunglm@monash.edu}

\author{Chakkrit Tantithamthavorn}
\affiliation{%
  \institution{Monash University}
  \city{Clayton}
  \country{Australia}}
\email{chakkrit@monash.edu}

\author{Michael Fu}
\affiliation{%
  \institution{Monash University}
  \city{Clayton}
  \country{Australia}}
\email{yeh.fu@monash.edu}

\author{John Grundy}
\affiliation{%
  \institution{Monash University}
  \city{Clayton}
  \country{Australia}}
\email{john.grundy@monash.edu}

\author{Hung Nguyen}
\affiliation{%
  \institution{Adelaide University}
  \city{Adelaide}
  \country{Australia}}
\email{hung.nguyen@adelaide.edu.au}

\author{Seyit Camtepe}
\affiliation{%
  \institution{CSIRO Data61}
  \city{Sydney}
  \country{Australia}}
\email{seyit.camtepe@data61.csiro.au}

\author{Paul Quirk}
\affiliation{%
  \institution{Defence Science and Technology Group}
  \city{Adelaide}
  \country{Australia}}
\email{paul.quirk@defence.gov.au}

\author{Dinh Phung}
\affiliation{%
  \institution{Monash University}
  \city{Clayton}
  \country{Australia}}
\email{dinh.phung@monash.edu}

\renewcommand{\shortauthors}{Van Nguyen et al.}

\begin{abstract}
Software vulnerabilities are a serious and crucial concern. Typically, in a program or function consisting of hundreds or thousands of source code statements, there are only a few statements causing the corresponding vulnerabilities.
Most current approaches to vulnerability labelling are done on a function or program level by experts with the assistance of machine learning tools. Extending this approach to the code statement
level is much more costly and time-consuming and remains an open problem. In this paper, we propose a novel end-to-end deep learning-based approach to identify the vulnerability-relevant
code statements of a specific function. Inspired by the specific structures observed in real-world vulnerable code, we first leverage mutual information for learning a set
of latent variables representing the relevance of the source
code statements to the corresponding function's vulnerability. We
then propose novel clustered spatial contrastive learning in order
to further improve the representation learning and the robust selection process of vulnerability-relevant code statements. Experimental results on real-world datasets of 200k+ C/C++ functions show the superiority of our method over other state-of-the-art baselines.  In general, our method obtains a higher performance in VCP, VCA, and Top-10 ACC measures of between 3\% to 14\% over the baselines when running on real-world datasets in an unsupervised setting. Our released source code samples are publicly available at \href{https://github.com/vannguyennd/livuitcl}{https://github.com/vannguyennd/livuitcl.}
\end{abstract}
\maketitle

\section{Introduction}
\label{sec:introduction}


It is common for 
computer software to contain 
software vulnerabilities (SVs), specific potential flaws, glitches,
weaknesses or oversights, that can be exploited by hackers or vandals
resulting in severe and serious economic damage \cite{Dowd2006}.
Potential vulnerabilities lurking in software development and deployment
processes can and do create severe security breaches, leading to a total
financial loss of  over \emph{USD} \emph{1 trillion}, a more
than 50 percent increase from 2018 \cite{mcafee_2020}. Modern computer software contains many thousands or even millions of lines of code and often follows diverse development processes that integrate code from different development teams. Thus finding such vulnerabilities is extremely challenging. Although
 many solutions have been proposed
for software vulnerability detection (SVD), the number of SVs and
the severity of the threats imposed by them have steadily increased
and caused considerable damage to individuals and companies \cite{ghaffarian2017software}.
These threats create an urgent need for more effective automatic tools and methods
to deal with a large amount of vulnerable code with a
minimal level of human intervention.

There have been many methods proposed for SVD based on rule-based, machine
learning or deep learning approaches. Most previous work in software
vulnerability detection \cite{shin2011evaluating,yamaguchi2011vulnerability,Almorsy_John,Li2016:VAV,Grieco2016,KimWLO17}
belongs to the former approaches and  require the knowledge of domain experts that
can be outdated and biased \cite{Zimmermann2009}. To mitigate this
problem, deep learning solutions have been used to conduct SVD and
have shown great advantages over machine learning approaches that use hand-crafted features, notably \cite{VulDeePecker2018,jun_2018,Dam2018,Li2018SySeVR,Duan2019,Cheng2019,Zhuang2020,nguyen2019deep,van-nguyen-dual-dan-2020,Zhen2021, ReGVD2021,fu2023chatgpt}. Despite achieving promising performance for SVD, current state-of-the-art deep learning-based methods (e.g., \cite{VulDeePecker2018,Dam2018,Li2018SySeVR,Zhen2021,ReGVD2021}) \emph{are only able to detect software vulnerabilities at a function or program level}. Particularly, these existing approaches detect whether a source code \textquotedblleft section\textquotedblright{} denoted by $F$ (e.g., a C/C++ function or program) is vulnerable. Hereafter, we use \textquotedblleft section\textquotedblright{}
or \textquotedblleft function\textquotedblright{} or \textquotedblleft program\textquotedblright{} to denote a collection of code statements. 


There have recently been some proposed approaches, notably
\cite{zhenli2020VulDeeLocator,van-ijcnn2021,IVDetect2021,LineVul2022,LineVD-2022,VELVET2022}, dealing with the \emph{statement-level vulnerability detection} problem. This includes highlighting statements that are highly relevant
to the corresponding function's vulnerability $Y$ (i.e., $Y\in\left\{ 0,1\right\} $
where \emph{1}: vulnerable and \emph{0}: non-vulnerable) and associated
code statements. Although these recently introduced methods achieve promising
results, they are limited in understanding and leveraging the relationships of hidden vulnerable patterns inside and between source code sections for facilitating statement-level vulnerability detection.

In this paper, we propose a novel end-to-end deep learning-based approach
for statement-level software vulnerability detection that allows us to
find and highlight code statements, in \emph{functions} or \emph{programs},
that are highly relevant to the presence of significant source code vulnerabilities. To determine key vulnerability-relevant code statements in each source code function $F$ consisting of $L$ code statements from $\boldsymbol{f}_{1}$ to $\boldsymbol{f}_{L}$ (\revise{in practice, each code statement is represented as a vector using a learnable embedding method, please refer to Section \ref{subsec:Data-Processing-and} for details}), we \textbf{(1) first introduce a \revise{learnable} selection process} $\varepsilon$ that aims to learn a set of independent Bernoulli latent variables $\mathbf{z}\in\{0,1\}^{L}$ representing the relevance of the code statements to the corresponding function's vulnerability $Y$, i.e., each element $z_{i}$ indicates whether $\boldsymbol{f}_{i}$ is related to the vulnerability of $F$. As $\mathbf{z}$ depends on $F$, we denote $\mathbf{z}(F)$. With $\mathbf{z}$, we construct $\tilde{F}=\varepsilon\left(F\right)$ (i.e., the subset of code statements that actually lead to the vulnerability $Y$ of the function $F$) by $\tilde{F}=\mathbf{z}(F)\odot F$, where $\odot$ represents the element-wise product. Inspired by \cite{learning-to-explain-l2x,van-ijcnn2021}, to ensure and enforce the selection process $\varepsilon$ obtaining the most meaningful $\tilde{F}$ (i.e. $\tilde{F}$ can predict the vulnerability $Y$ of $F$ correctly), we then \textbf{(2) maximize the mutual information between $\tilde{F}$ and $Y$}. 

From the real source code sections (i.e., functions or programs), we observe that there are often groups of several core statements that cause their vulnerability. If
we group these core statements together, we have vulnerability patterns
shared across vulnerable source code sections. 
These hidden vulnerability patterns
can be embedded into real-world source code sections at different
spatial locations to form realistic vulnerable source code sections.
Given a set of vulnerable source code sections,
we \textit{need to devise a  mechanism to guide the selection process}
$\varepsilon$ to select and highlight hidden vulnerability patterns.
This is a challenging task since vulnerability patterns are
hidden and can be embedded into real vulnerable source code sections
at different spatial locations. To this end, to characterize a
vulnerable source code section $F$, we consider $F^{top}$ that includes
$K$ statements in $F$ with the top $K$ highest selection probabilities.
We further observe that a vulnerability type consists of several vulnerability
patterns, and vulnerable source code sections originated from the same
vulnerability pattern possess very similar the top $K$ statements
$F^{top}$ which tend to form well-separated clusters. Based on this
observation, we \textbf{ (3) propose a clustered spatial contrastive learning term},
inspired by supervised contrastive learning \cite{Khosla2020},
which ensures and enforces three important properties in the source code data. This includes (i) the vulnerable and non-vulnerable source code sections should have different representations of $F^{top}$, (ii) the vulnerable source code sections from different hidden vulnerability patterns are also encouraged to have different representations of $F^{top}$ while (iii) the vulnerable source code sections in the similar hidden vulnerability patterns are trained to have close representations of $F^{top}$. Ensuring these properties facilitates the selection process for helping boost the data representation learning in figuring out and selecting the code vulnerable statements of vulnerable  source code sections.

\vspace{1mm}
The key contributions of this work include:
\begin{itemize}
\item We introduce an end-to-end deep learning-based approach for statement-level
vulnerability detection based on an information-theoretic
perspective in forming the model selection and guiding the whole training
process. We propose a novel clustered spatial contrastive learning
term inspired by contrastive learning \cite{Khosla2020}
to model important properties in the relationship of vulnerable patterns
between the source code sections. 
\revise{Our method shares the intuition with \cite{van-ijcnn2021} of using mutual information in selecting the important subset of code statements in each source code function, however, in our method, we propose a novel clustered spatial contrastive learning to learn important properties from the source code data for boosting the data representations and learning hidden vulnerability patterns that facilitate the selection of vulnerable code statements in vulnerable source code sections (i.e., functions or programs).}
\item We conduct extensive experiments on real-world source code datasets,
including CWE-119 and CWE-399 collected by \cite{VulDeePecker2018},
and a big C/C++ source code dataset, namely Big-Vul, provided by \cite{Bigdata2020}.
Our experiments on these three real-world datasets of 200k+ C/C++ functions show the superiority of our proposed method in selecting and highlighting the core vulnerable
statements over state-of-the-art baselines.
\end{itemize}

\section{Motivation}
%

\begin{figure}[ht]
\begin{centering}
\vspace{0mm}
\includegraphics[width=0.35\columnwidth]{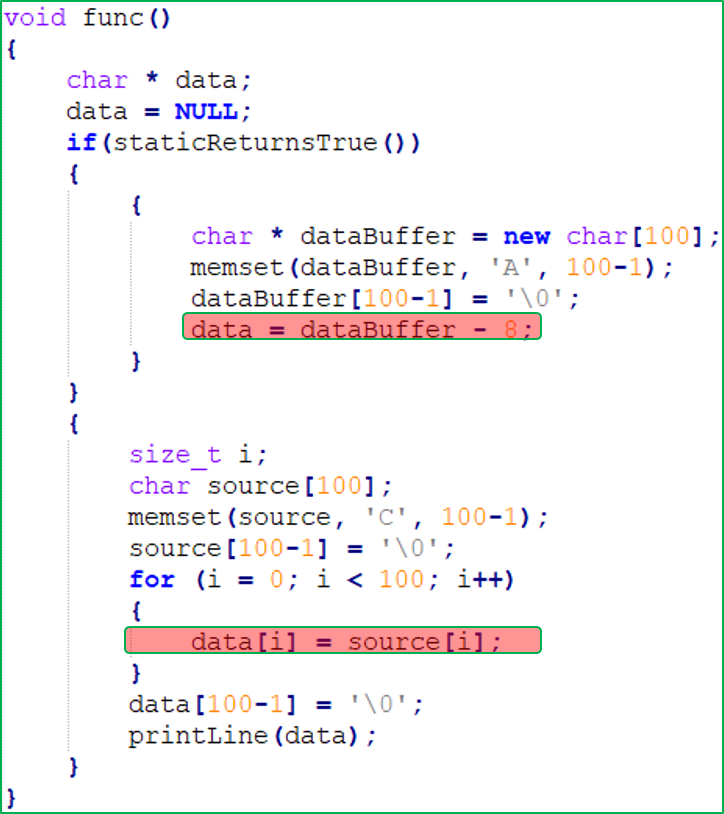}
\par\end{centering}\vspace{-3mm}
\caption{An example of a buffer error vulnerability source code function. \label{motivation_example}}
\end{figure}

Real-world source code programs or functions
can consist of many hundreds or thousands of source code statements, and
only a few of them (usually a few core statements) cause the corresponding
vulnerabilities. Figure \ref{motivation_example} shows an example of a simple vulnerable C/C++ source
code function. Among many lines of code statements, only two statements, highlighted in red, actually lead to the function's
vulnerability. The core statements underpinning a vulnerability are
even much sparser in the source code of real-world applications. In Figure \ref{motivation_example}, for
demonstration purposes, we choose a short C/C++ source code function in which
the statement \textit{\textquotedblleft data = dataBuffer - 8;\textquotedblright{}}
is a vulnerability because we set the data pointer to before the allocated
memory buffer, and the statement \textquotedblleft \textit{data{[}i{]}
= source{[}i{]};}\textquotedblright{} is a potential flaw due to possibly
copying data to memory before the destination buffer.

\vspace{-2mm}

It is worth noting that most of the existing state-of-the-art SVD approaches can only detect whether a source code \textquotedblleft section\textquotedblright{} $F$ (e.g., a C/C++ function or program) is vulnerable. In contrast, by doing statement-level vulnerability detection (i.e., learning to select the vulnerable statements in each vulnerable source code section), e.g., \cite{zhenli2020VulDeeLocator,van-ijcnn2021,IVDetect2021,LineVul2022,LineVD-2022,VELVET2022}, we can help speed up the process of isolating
and detecting software vulnerabilities thereby reducing the time and cost involved. Doing statement-level vulnerability detection is the main target of our proposed approach.

\begin{figure*}[ht]%
\begin{centering}
\begin{tabular}{c}
\includegraphics[width=0.9\textwidth]{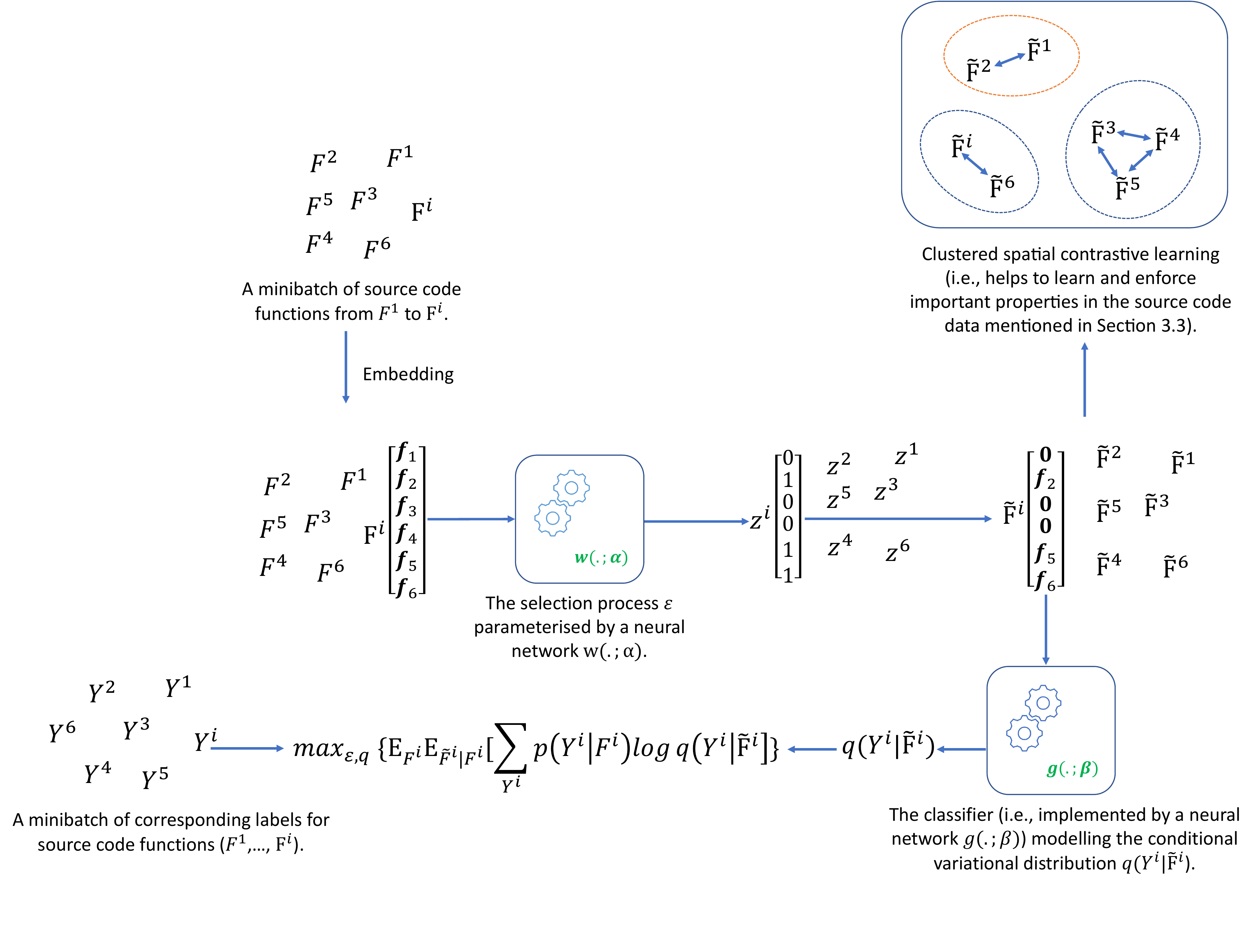}\tabularnewline
\end{tabular}\vspace{-10mm}
\par\end{centering}
\caption{The overall architecture of our LEAP method. Given a mini-batch of source code sections (i.e., from $F^{1}$ to $F^{i}$), to each code section, e.g., $F^{i}$, the selection process $\varepsilon$ learns a set of independent Bernoulli latent variables $\mathbf{z}\in\{0,1\}^{L}$ representing the relevance of the code statements to the corresponding function's vulnerability $Y^{i}$. For demonstration purposes, we assume that there are six code statements in $F^{i}$. However, in reality, this number can be in the hundreds. We then construct $\tilde{F}^{i}$ (i.e., the subset of code statements that actually lead to the vulnerability $Y^{i}$) by  $\tilde{F}^{i}=\mathbf{z}^{i}(F^{i})\odot F^{i}$. Importantly, for ensuring and enforcing $\varepsilon$ obtaining the most meaningful $\tilde{F}^{i}$ (i.e., $\tilde{F}^{i}$ can predict the vulnerability $Y^{i}$ of $F^{i}$ correctly), we maximize the mutual information between $\tilde{F}^{i}$ and $Y^{i}$. The proposed clustered spatial contrastive learning helps to learn and enforce important properties in the source code data for boosting the data representation learning in figuring out and selecting vulnerable patterns and vulnerable statements in each vulnerable source code section.\label{fig:model-architecture-leap}}
\vspace{-5mm}
\end{figure*}%

\section{Our Approach}
\label{sec:approach}


\subsection{The problem setting}


We denote a source code section (e.g., a C/C++ function or program)
as $F=[\boldsymbol{f}_{1},\dots,\boldsymbol{f}_{L}]$, which consists
of $L$ code statements from $\boldsymbol{f}_{1}$ to $\boldsymbol{f}_{L}$
($L$ can be a large number, e.g., hundreds or thousands). \textit{In practice, each code statement is represented as a vector, which is extracted by some embedding methods. As those embedding methods are not the focus of this paper, we leave these details to the experiment section.}
We assume that $F$'s vulnerability $Y\in\left\{ 0,1\right\} $ (where
\emph{1}: vulnerable and \emph{0}: non-vulnerable) is observed (labeled
by experts). As previously discussed, there is usually a small subset
with $K$ code statements that actually lead to $F$ being vulnerable,
denoted as $\tilde{F}=[\boldsymbol{f}_{i_{1}},\dots,\boldsymbol{f}_{i_{K}}]=[\boldsymbol{f}_{j}]_{j\in S}$
where $S=\{i_{1},\dots,i_{K}\}\subset\{1,\dots,L\}$ ($i_{1}<i_{2}<...<i_{K}$).
To select the vulnerability-relevant statements $\tilde{F}$ for each
specific source code section $F$, we apply to use a learnable
selection process $\varepsilon$, i.e., $\tilde{F}=\varepsilon\left(F\right)$,
whose training principle and construction are presented in the following subsection. We then propose novel clustered
spatial contrastive learning, which can model important properties
for the relationship of the source code sections, presented in the following sections to further improve the
representation learning and the robust selection process of $\tilde{F}$.

The focus of our proposed method is to tackle the most challenging reality setting where most available datasets only have the vulnerability
label (i.e., $Y$) at the source code function level i.e., only denote whether
a function $F$ is vulnerable, by experts with the assistance of machine learning
tools. They do not contain information
of which particular source code statement(s) cause vulnerabilities. In the training
process, our method only requires a vulnerability label
at the source code function level (i.e., $Y$) and is capable of pointing
out the vulnerability-relevant statements. \emph{In the context of statement-level vulnerability detection}, this setting is considered as \emph{the unsupervised one} as mentioned in \cite{van-ijcnn2021},
meaning that the training process does not require labels at the code
statement level (i.e., the ground truth of vulnerable code statements
causing vulnerabilities). The ground truth of vulnerable code statements
causing vulnerabilities in source code sections is only used in the evaluation process.

\revise{\textbf{In Sections \ref{subsec:Training-principle-and} and \ref{subsec:Clustered-spatial-contrastive} as follows,}} we present the methodology of our proposed approach for identifying source code statements causing the vulnerability of each source code section. \textbf{In particular, in Section \ref{subsec:Training-principle-and}}, we first present the way we construct a learnable selection process $\varepsilon$ and use mutual information to ensure and enforce the selection process $\varepsilon$ obtaining the most important source code statements, denoted by $\tilde{F}$, which causes the corresponding label $Y$ as well as forms the corresponding vulnerability pattern of the source code section $F$. \textbf{In Section \ref{subsec:Clustered-spatial-contrastive}}, we then present our novel clustered spatial contrastive learning in guiding the selection process $\varepsilon$ to model important properties 
in the relationships of vulnerable patterns between the source code sections to boost the data representation learning in figuring out and selecting the source code vulnerable statements.

\subsection{Learning to select vulnerable-relevant statements and the training principle\label{subsec:Training-principle-and}}

\subsubsection{\revise{\textbf{Vulnerability-relevant statements selection process}}}


Giving a source code function $F$ consisting of $L$ code statements from $\boldsymbol{f}_{1}$ to $\boldsymbol{f}_{L}$ (i.e., \revise{in practice, each code statement $\boldsymbol{f}_{i}$ is represented as a vector using a learnable embedding method, please refer to Section \ref{subsec:Data-Processing-and} for details}), to figure out its vulnerability-relevant code statements $\tilde{F}$, we introduce a learnable selection process $\varepsilon$ \revise{(i.e., $\mathbb{R}^L \mapsto [0,1]^L$)} aiming to learn a set of independent Bernoulli latent variables $\mathbf{z}\in\{0,1\}^{L}$ representing the relevance (importance) of the code statements to the corresponding function's vulnerability $Y$. Specifically, each element $z_{i}$ in $\mathbf{z}=\{z_{i}\}_{i=1}^{L}$ indicates whether $\boldsymbol{f}_{i}$ is related to the vulnerability $Y$ of $F$ (i.e., if  $z_{i}$ is equal to $1$, the statement $\boldsymbol{f}_{i}$ plays an important role causing the vulnerability $Y$). 

\revise{To} construct $\mathbf{z}=\{z_{i}\}_{i=1}^{L}$, we model $\mathbf{z} \sim \mathrm{MultiBernoulli}(\bm{p}) = \prod_{i=1}^{L}\text{Bernoulli}(p_{i})$,
which indicates $\boldsymbol{f}_{i}$ is related to the vulnerability $Y$ of $F$ with probability $p_{i}$ \revise{where}
$p_{i}=\omega_{i}\left(F;\alpha\right)$ \revise{with} $\omega$ is a neural
network parameterized by parameter $\alpha$. The neural network $\omega$ takes $F$ as input and outputs corresponding $\vp=\{p_{i}\}_{i=1}^{L}$. 

As $\mathbf{z}$ depends on $F$, we denote $\mathbf{z}(F)$. With $\mathbf{z}$, we construct $\tilde{F}=\varepsilon\left(F\right)$ (i.e., the subset of code statements that actually lead to the vulnerability $Y$ of the function $F$) by $\tilde{F}=\mathbf{z}(F)\odot F$, where $\odot$ represents the element-wise product. \revise{Please refer to 
Figure \ref{fig:model-architecture-leap} for the corresponding visualization.}

\subsubsection{\revise{\textbf{Gumbel-Softmax trick for continuous optimization}}}

Recall that the vulnerability-relevant selected code statements $\tilde{F}$ are constructed by $\tilde{F}=\mathbf{z}(F)\odot F$ where $\mathbf{z} \sim \mathrm{MultiBernoulli}(\bm{p})$ with $\vp$ is the output of the selection process $\varepsilon$ parameterised by a neural network  $\omega(.,\alpha)$. \revise{
To make this computational process (having a sampling operation) continuous and differentiable during training, we apply the temperature-dependent Gumbel-Softmax trick {\cite{jang2016categorical}} for relaxing each Bernoulli variable $z_{i}$.} Particularly, we sample $z_{i}\left(F;\alpha\right)\sim\text{Concrete}(\omega_{i}(F;\alpha),1-\omega_{i}(F;\alpha))$:
\[z_{i}\left(F;\alpha\right)=\frac{\exp\{(\log\omega_{i}\left(F;\alpha\right)+a_{i})/\nu\}}{\exp\{(\log\omega_{i}\left(F;\alpha\right)+a_{i})/\nu\}+\exp\{(\log\left(1-\omega_{i}\left(F;\alpha\right)\right)+b_{i})/\nu\}}
\]

\revise{where $\nu$ is a temperature parameter (i.e., that allows us to control how closely a continuous representation from a Gumbel-Softmax distribution approximates this from the corresponding discrete representation from a discrete distribution (e.g., the Bernoulli distribution), random noises $a_{i}$ and $b_{i}$ independently drawn from \textbf{Gumbel} distribution
$G = - \log(- \log u)$ with $\ u \sim \textbf{Uniform}(0,1)$.}

\subsubsection{\revise{\textbf{Mutual information for guiding the selection process}}}
\revise{
\paragraph{Mutual information} In information theory \cite{TheoryofC, EofIT2006}, mutual information (MI) is used to measure the mutual dependence between two random variables. In particular, MI quantifies the amount of information obtained about one random variable by observing the other random variable. To illustrate, consider a scenario where A denotes the outcome of rolling a standard 6-sided die, and B represents whether the roll results in an even number (0 for even, 1 for odd). Evidently, the information conveyed by B provides insights into the value of A, and vice versa. In other words, these random variables possess mutual information.

\paragraph{How mutual information guides the selection process}
Leveraging the important property of mutual information (quantifying the amount of information obtained about one random variable by observing the other random variable) and inspired by \cite{learning-to-explain-l2x}, we maximize the mutual information between $\tilde{F}$ and $Y$ as mentioned in Eq. (\ref{eq:max_info}). Specifically, we aim that via Eq. (\ref{eq:max_info}), by using the information from $Y$, the selection process $\varepsilon$ will be enforced to obtain the most meaningful $\tilde{F}$ (i.e., $\tilde{F}$ can predict the vulnerability $Y$ of $F$ correctly).
}
If we view $\tilde{F}$ and $Y$ as random variables,
\textit{the selection
process $\varepsilon$ is learned by maximizing the mutual information
between $\tilde{F}$ and $Y$} as follows:
\vspace{-0mm}
{\small{}
\begin{equation}
\max_{\varepsilon}\,\mathbb{I}(\tilde{F},Y).\label{eq:max_info}
\end{equation}
}{\small\par}
\vspace{-0mm}
We expand Eq. (\ref{eq:max_info}) further as the Kullback-Leibler
divergence \revise{(i.e., it measures the relative entropy or difference in information represented by two distributions)} of the product of marginal distributions of $\tilde{F}$
and $Y$ from their joint distribution:
{\small{}
\begin{align}
\mathbb{I}(\tilde{F},Y)= & \int p(\tilde{F},Y)\log\frac{p(\tilde{F},Y)}{p(\tilde{F})p(Y)}d\tilde{F}dY\nonumber \\ 
\geq & \int p(Y,\tilde{F})\log\frac{q(Y\mid\tilde{F})}{p(Y)}dYd\tilde{F}\label{eq:1sm}
\end{align}
}{\small\par}

\revise{ In practice, estimating mutual information is challenging as we typically have access to samples but not the underlying distributions. Therefore,} in the above derivation, we use a variational distribution
$q(Y|\tilde{F})$ to approximate the posterior $p(Y\mid\tilde{F})$,
hence deriving a variational lower bound of $\mathbb{I}(\tilde{F},Y)$
for which the equality holds if $q(Y\mid\tilde{F})=p(Y\mid\tilde{F})$.
This can be further expanded as:

{\small{}
\begin{align}
\mathbb{I}(\tilde{F},Y)\geq & \int p(Y,\tilde{F},F)\log\frac{q(Y\mid\tilde{F})}{p(Y)}dYd\tilde{F}dF\nonumber \\
= & \mathbb{E}_{F}\mathbb{E}_{\tilde{F}|F}[\sum_{Y}p(Y|F)\log q(Y|\tilde{F})]+\text{const}\label{eq:2sm}
\end{align}
}{\small\par}

We note that $\tilde{F}|F:=\tilde{F}\sim p(\cdot|F):=\varepsilon\left(F\right)$
is the same representation of the selection process and $p(Y|F)$
as mentioned before is the ground-truth conditional distribution
of the $F$'s label on all of its features. 

To model the conditional variational distribution $q(Y|\tilde{F})$,
we introduce a classifier implemented with a neural network $g(\tilde{F};\beta)$, which
takes $\tilde{F}$ as input and outputs its corresponding label. Our objective is to learn the selection process as well as the classifier
to maximize the mutual information:

{\small{}
\begin{equation}
\text{max}_{\varepsilon,q}\{\mathbb{E}_{F}\mathbb{E}_{\tilde{F}|F}[\sum_{Y}p(Y|F)\log q(Y|\tilde{F})]\}.\label{eq:max_info_mutual}
\end{equation}
}{\small\par}

The mutual information facilitates a joint training process for the
classifier and the selection process. \textit{The classifier learns to
identify a subset of features leading to a data sample's label
while the selection process is designed to select the best subset
according to the feedback of the classifier.}

\subsection{Clustered spatial contrastive learning\label{subsec:Clustered-spatial-contrastive}}

\paragraph{\textbf{Motivation.}}

\begin{figure}[ht]
\begin{centering}
\includegraphics[width=0.55\textwidth]{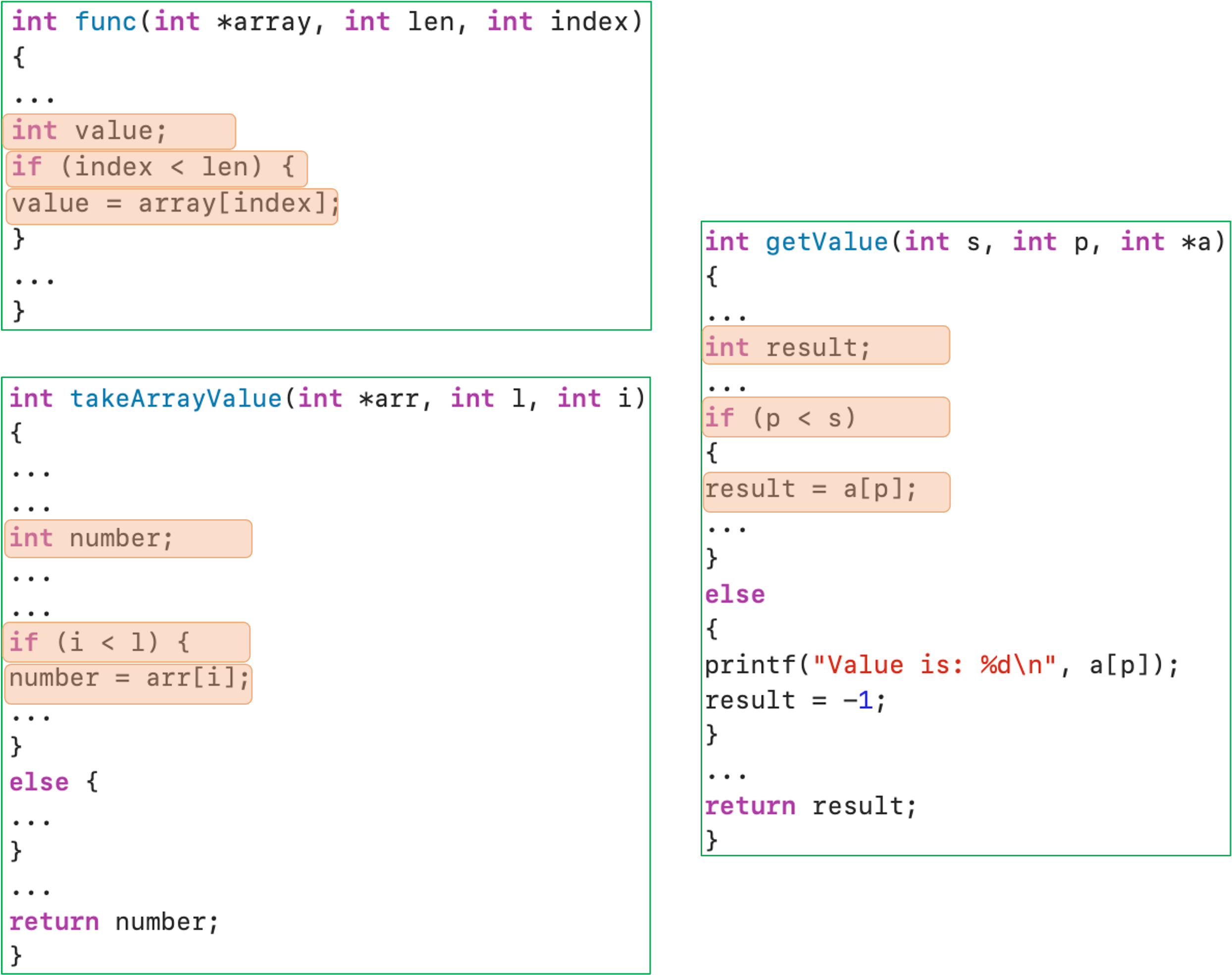}
\par\end{centering}
\vspace{-1mm}
\caption{An example of the\emph{ improper validation of array index} flaw pattern
(i.e., the top-left-hand figure) with two real-world source code functions
(i.e., \emph{takeArrayValue} and \emph{getValue}) containing this pattern.
In each function, there are some parts (i.e., denoted by \emph{\textquotedblleft ...\textquotedblright}) omitted for brevity.\label{fig:Buffer-Copy-improper-movitation-1}}
\vspace{-1mm}
\end{figure}

For each vulnerable function $F$, we observe that there are few statements
causing a vulnerability. If we group those core statements together,
they form \emph{vulnerability patterns}. For example, the top-left-hand figure in Figure
\ref{fig:Buffer-Copy-improper-movitation-1} shows a vulnerability
pattern named the \emph{improper validation of array index} flaw pattern
for the \emph{buffer overflow error} in which the software performs
operations on a memory buffer, but it can read from or write to a
memory location that is outside of the intended boundary of the buffer.
More specifically, this aims to get a value from an array (i.e., \emph{int
{*}array}) via specific \emph{index} and save this value into a variable
(i.e., \emph{value}). However, this only verifies that the given array
\emph{index} is less than the maximum length of the array using the
statement "\emph{if(index<len)}" but does not check for the minimum
value, hence allowing a negative value to be accepted as the input
array index, which will result in an out-of-bounds read and may allow
access to sensitive memory.

As shown in Figure \ref{fig:Buffer-Copy-improper-movitation-1}, this
vulnerability pattern is embedded into real-world functions \emph{getValue}
and \emph{takeArrayValue} in which the core statements in the vulnerability
pattern are placed into different spatial locations under different
variable names. We wish to guide the selection process so that the
vulnerable source code sections originated from the same vulnerability
pattern have similar selected and highlighted statements which commonly
specify this vulnerability pattern. This is challenging because the
common vulnerability pattern is embedded into those source code sections
at different spatial locations. To address this issue, given a source
code section $F$, we define $F^{top}$ as a subset of $F$ including
its $K$ statements with the top $K$ selection probability $p_{i}=\omega_{i}\left(F;\alpha\right)$
and employ $F^{top}$ to characterize the predicted vulnerability
pattern of $F$. It is worth noting that the statements $F^{top}$
preserves the order in $F$. 

\begin{figure*}[ht]%
\begin{centering}
\begin{tabular}{c}
\includegraphics[width=0.9\textwidth]{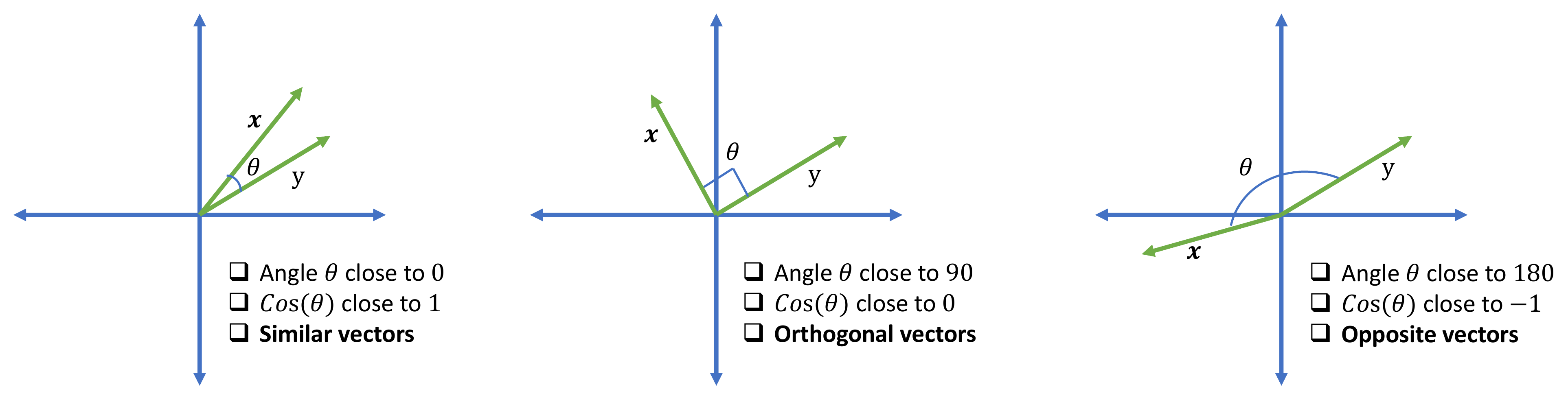}\tabularnewline
\end{tabular}\vspace{-1mm}
\par\end{centering}
\caption{\revise{A graphic showing two vectors with cosine similarities close to 1, close to 0, and close to -1. The similarity of two vectors is measured by the cosine of the angle between them. The similarity can take values between -1 and +1. Smaller angles between vectors produce larger cosine values, indicating greater cosine similarity.}\label{fig:cosine-similarity}}
\vspace{-1mm}
\end{figure*}%

To enforce two vulnerable source code sections originated from the
same vulnerability pattern having the same $F^{top}$, a possible
solution is to employ the supervised contrastive learning \cite{Khosla2020}
to reach the following objective function based on the contrastive
learning principle as follows:
{\small{}
\begin{align}
\mathcal{L}_{scl} = \sum_{i\in I}1_{Y^{i}=1}\frac{-1}{\left|P(i)\right|}\sum_{p\in P(i)}\log\frac{\exp(\text{sim}({F^{i}}^{top},{F^{p}}^{top})/\tau)}{\sum\limits_{a\in A(i)}\exp(\text{sim}({F^{i}}^{top},{F^{a}}^{top})/\tau)}\label{eq:cl_spatial_cl}
\end{align}
}{\small\par}

where $I\equiv\{1...m\}$ is a set of indices of input data in a specific
mini-batch; $\text{sim}$ is the cosine similarity \revise{(\textit{i.e., a metric used to measure the similarity of two vectors. Specifically, it measures the similarity in the direction or orientation of the vectors ignoring differences in their magnitude or scale. The similarity of two vectors is measured by the cosine of the angle between them (please refer to Figure \ref{fig:cosine-similarity} for details). Note that to form the corresponding vector for each $F^{top}$ of each source code section $F$ in the latent space for calculating the cosine similarity, we simply concatenate all vectors where each vector stands for a representation of a code statement in $F^{top}$})}; $\tau>0$ is a
scalar temperature parameter; $A(i)\equiv I\setminus\{i\}$; $P(i)\equiv\{p\in A(i):Y^{p}=1\}$
is the set of indices of vulnerable source code sections with the
label $1$ ($1:$\emph{vulnerable} and $0$: \emph{non-vulnerable})
in the mini-batch except $i$;  $\left|P(i)\right|$ is its cardinality;
and $1_{A}$ represents the indicator function.

It can be observed that although the objective function in (\ref{eq:cl_spatial_cl})
encourages vulnerable source code sections to share the same selected
and highlighted vulnerability pattern, it seems to overdo this by
forcing all vulnerable source code sections to share the same vulnerability
pattern. \textbf{In what follows, we present an efficient workaround to mitigate
this drawback.}

\paragraph{\textbf{Clustered spatial contrastive learning.}\label{par:Clustered-spatial-contrastive}}

\begin{figure}[ht]
\begin{centering}
\includegraphics[width=0.8\columnwidth]{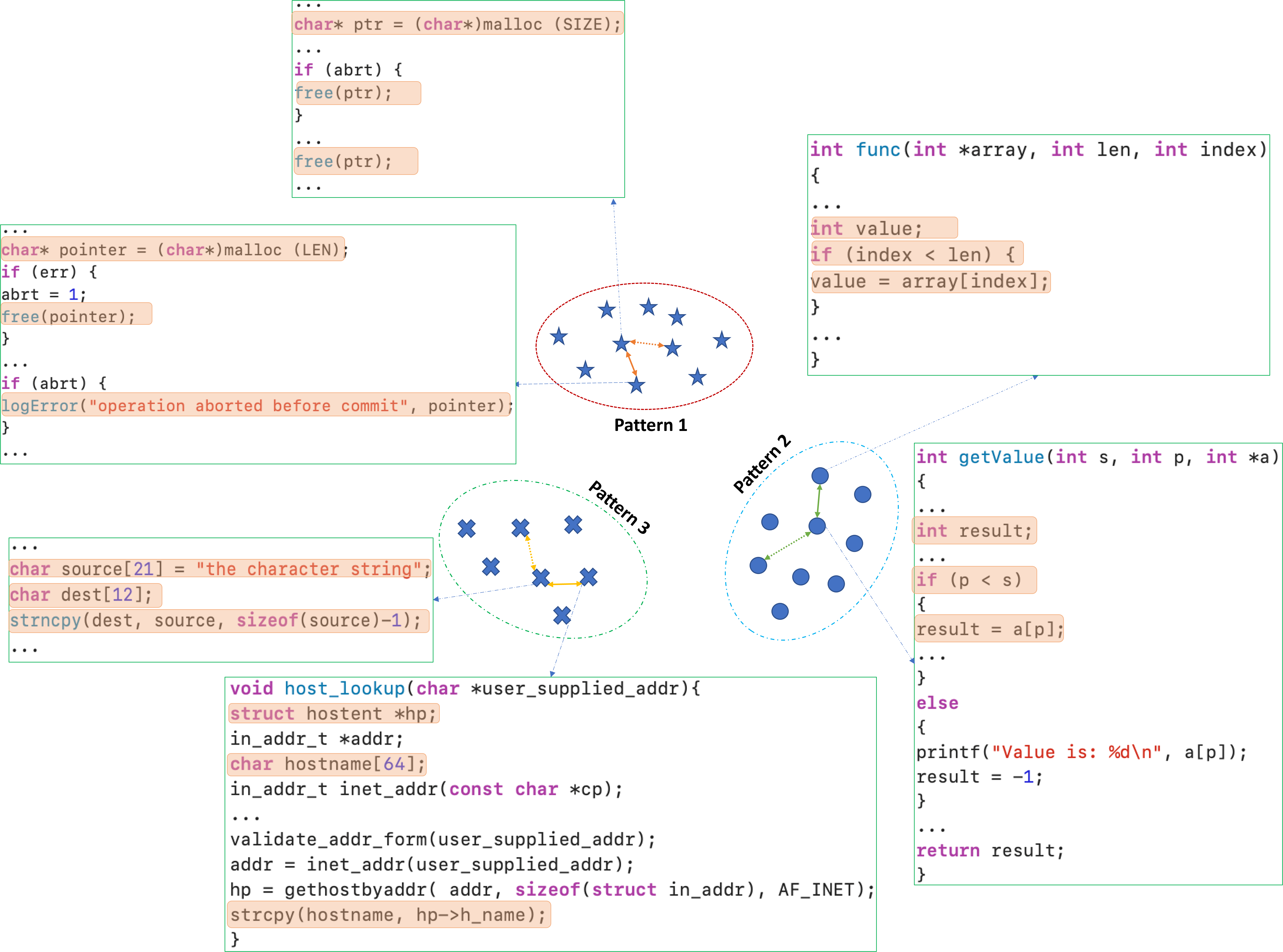}
\par\end{centering}\vspace{0mm}
\caption{A demonstration of different vulnerability patterns forming different patterns in the latent space for the buffer overflow error. In particular, Pattern 1 stands for the expired pointer dereference flaw \textit{in which the program dereferences
a pointer containing a location for memory that was previously valid, but it is no longer valid.} Pattern 2 represents the improper validation of the array index flaw \textit{in which the product uses untrusted input when using an array index, but the
product does not validate or incorrectly validates the index
to ensure the index references a valid position within the array} while Pattern 3 presents the buffer access with an incorrect length value flaw \textit{in which the software uses a sequential operation
to read or write a buffer, but it may use an incorrect length
value resulting in accessing memory that is outside of the
bounds of the buffer.} 
Note that each
data point in a pattern is a specific $F^{top}$ (e.g., the colored background lines) of a corresponding
function $F$. In this demonstration, we assume that there are three different patterns causing the buffer overflow error. In reality, the number of vulnerability
patterns causing the buffer overflow error can be higher. \label{fig:Cluster_spatial_examples-1}}
\vspace{0mm}
\end{figure}

We observe that each different vulnerability type might have some
different vulnerability patterns causing it. For example, the\emph{
buffer overflow error} can have \emph{"buffer access with incorrect length}", "\emph{improper validation of array index}", or "\emph{expired pointer dereference}" as mentioned in Figure
\ref{fig:Cluster_spatial_examples-1}.
We further observe that the vulnerable source code
sections originated from the same vulnerability pattern tend to have the similar
$F^{top}$ and form a well-separated cluster as shown in Figure
\ref{fig:Cluster_spatial_examples-1}. 
\vspace{0mm}

Therefore, we propose to do
clustering analysis (e.g., $k$-means) on $F^{top}$ to group vulnerable
source code sections with the same vulnerability patterns and employs
contrastive learning to force them to become more similar as follows:

{\small{}
\begin{align}
\mathcal{L}_{cscl} = \sum_{i\in I}1_{Y^{i}=1}\frac{-1}{\left|C(i)\right|}\sum_{c\in C(i)}\log\frac{\exp(\text{sim}({F^{i}}^{top},{F^{c}}^{top})/\tau)}{\sum\limits_{a\in A(i)}\exp(\text{sim}({F^{i}}^{top},{F^{a}}^{top})/\tau)} \label{eq:cluster_spatial_cl}
\end{align}
}{\small\par}

where $I\equiv\{1...m\}$ is a set of indices of input data in a specific
mini-batch; $A(i)\equiv I\setminus\{i\}$, $C(i)\equiv\{c\in A(i):\bar{Y}^{c}=\bar{Y}^{i}\,\text{and}\,Y^{c}=1\}$
is the set of indices of vulnerable source code sections labeled $1$
which are in the same cluster as $F^{i}$ except $i$; and $\left|C(i)\right|$
is its cardinality. \emph{} Note that in Eq. (\ref{eq:cluster_spatial_cl}),
we apply $k$-means for the current mini-batch and denote $\bar{Y}^{i}$
as the cluster label of the source code section $F^{i}$.

It is worth noting that our proposed clustered spatial contrastive learning helps to \textbf{enforce three important properties in the source code data including} (i) the vulnerable and non-vulnerable source code sections should have different representations of $F^{top}$, (ii) the vulnerable source code sections from different hidden vulnerability patterns are also encouraged to have different representations of $F^{top}$ while (iii) the vulnerable source code sections in the similar hidden vulnerability patterns are trained to have close representations of $F^{top}$. Ensuring these properties facilitates the selection process for helping boost the data representation learning in figuring out and selecting the source code vulnerable statements of vulnerable source code sections (e.g., functions).

Combining the objective functions in Eqs. (\ref{eq:max_info_mutual}
and \ref{eq:cluster_spatial_cl}), we arrive at the following objective
function:
\begin{equation}
\text{max}_{\varepsilon,q}\{\mathbb{E}_{F}\mathbb{E}_{\tilde{F}|F}[\sum_{Y}p(Y|F)\log q(Y|\tilde{F})]-\alpha\mathcal{L}_{cscl}\}\label{eq:final-objective-function}
\end{equation}

where $\alpha>0$ is the trade-off hyper-parameter. The overall architecture of our proposed method is depicted in Figure \ref{fig:model-architecture-leap}.
\vspace{2mm}
\section{Experimental Design}
\label{sec:exp_design}

\subsection{Research Questions}
We wanted to evaluate our \ourapp~approach and compare it with other baselines for statement-level vulnerability detection. 
Below, we present the key research questions to answer in our study.

\vspace{1mm}
\textbf{(RQ1) \rqone}
This is the main focus of our proposed method where we aim to tackle the most challenging reality setting in which most publicly available datasets only have the vulnerability
label (i.e., $Y$) at the source code function level, i.e., only denote whether a function $F$ is vulnerable, by experts with the assistance of machine learning tools. They do not contain information
on which source code statement(s) cause vulnerabilities. Therefore, proposing a method for statement-level vulnerability detection that can be trained without using any ground truth of vulnerable code statements play an important role in this challenging reality context.

\vspace{1mm}
\textbf{(RQ2) \rqtwo}
In some scenarios, a small portion of the training data may have the ground truth of vulnerable code statements (i.e., the semi-supervised setting). Therefore, in this research question, we aim to investigate the performance of our proposed method (and the baselines) when we incorporate the annotations of the vulnerable statements of a small portion of the training data in the training process. In particular, we assume that 10\% of the training data has the ground truth of vulnerable code statements. 

\vspace{1mm}
\textbf{(RQ3) \rqthree}
We aim to investigate the effectiveness of our proposed cluster spatial contrastive learning in Eq. (\ref{eq:cluster_spatial_cl}) in terms of facilitating the data representation learning process and helping figure out the hidden vulnerability pattern for improving the model's ability to select the vulnerable statements in each source code data compared to the cases of using the normal contrastive learning in Eq.(\ref{eq:cl_spatial_cl}) and without using contrastive learning (i.e., in this case, we only use the mutual information for guiding the whole training process).

\subsection{Studied datasets}
We used three real-world datasets including \textit{the CWE-399 dataset} with 1,010 and  1,313 vulnerable/non-vulnerable functions for resource management
error vulnerabilities, \textit{the CWE-119 dataset} with 5,582 and 5,099 vulnerable/non-vulnerable functions
for the buffer error vulnerabilities, and \textit{a big C/C++
dataset}, namely Big-Vul, provided by \cite{Bigdata2020} containing many types
of vulnerabilities such as Out-of-bounds Write, Improper Input Validation,
and Path Traversal. 

For the CWE-399 and CWE-199 datasets collected by \cite{VulDeePecker2018}, we used the ones processed by \cite{van-ijcnn2021}. Additionally, the Big-Vul dataset, one of the largest vulnerability datasets having the ground truth at the statement level, was collected from 348 open-source Github projects from 2002 to 2019.
It consists of 188,636 C/C++ source code functions where a ratio of vulnerable functions is 5.7\% (i.e., 10,900 vulnerable functions).

For the training process of the unsupervised setting, we do not use the information of vulnerable statements (i.e., the vulnerability labels at the statement level). However, this information is necessary to evaluate the models' performance.

\subsection{Data processing and embedding\label{subsec:Data-Processing-and}}
\revise{We preprocessed the datasets before injecting them into deep neural networks.
In particular, we standardized the source code by removing comments and non-ASCII characters for the Big-Vul dataset while for the CWE-399 and CWE-199 datasets 
, we used the ones preprocessed by \cite{van-ijcnn2021}.
We then embedded source code statements into vectors. For instance, consider the following
statement (C programming language) \emph{"for(i=0;i<10;i++)}", to embed this code statement, we tokenized it to a sequence of tokens (e.g., \emph{for,(,i,=,0,;,i,<,10,;,i,++,)}), and then we used a 150-dimensional token embedding followed by a Dropout layer with a dropped fixed probability $p=0.2$ and (a 1D convolutional layer with the filter size $150$ and kernel size $3$, and a 1D max pooling layer) or (a 1D max pooling layer) to encode each statement in a function $F$.\emph{ }Finally, a mini-batch of functions in which each function consisting of $L$ encoded statements was fed to the models.
}

\revise{
Different from the L2X \cite{learning-to-explain-l2x}, ICVH \cite{van-ijcnn2021}, and our \ourapp~ methods, the LineVul method \cite{LineVul2022}
based on the CodeBERT tokenizer and CodeBERT pre-trained model from
\cite{CodeBERT2020} to tokenize and generate vector representations
of source code functions.}

\subsection{Measures and evaluation\label{subsec:Measures-and-Evaluation}}
The main purpose of our \ourapp~method is to support security analysts
and code developers to narrow down the vulnerable scope in their search for
vulnerable statements. This would be helpful in the context that they
need to identify several vulnerable statements from hundreds or thousands
of lines of code. We aim to identify source code statements (e.g.,\emph{
top K=}10) so that with a high probability these statements cover most
or all vulnerable statements in the corresponding source code section. 

To evaluate the performance of our
\ourapp~method and baselines, we use two main measures introduced
in \cite{van-ijcnn2021} including: \emph{vulnerability coverage
proportion (VCP)} (i.e., the proportion of correctly detected vulnerable
statements over all vulnerable statements in a dataset) and \emph{vulnerability
coverage accuracy (VCA)} (i.e., the ratio of the successfully detected
functions, having all vulnerable statements successfully detected,
over all functions). 

In addition to the VCP and VCA measures, we also report two other
measures including \emph{Top-10 Accuracy} (i.e., it measures the percentage
of vulnerable functions where at least one actual vulnerable lines
appear in the top-10 ranking) and Initial False Alarm (\emph{IFA})
(i.e., it measures the number of incorrectly predicted lines (i.e.,
non-vulnerable lines incorrectly predicted as vulnerable or false
alarms) that security analysts need to inspect until finding the first
actual vulnerable line for a given function) used in \cite{LineVul2022}.

\subsection{Baseline Methods}
We used several baseline approaches to compare to our
proposed method \ourapp~(Statement-Level Vulnerability Detection: \underline{Lea}rning Vulnerability \underline{P}atterns Through Information Theory and Contrastive Learning). The main baselines to our method are \textbf{ICVH} \cite{van-ijcnn2021} and \textbf{LineVul} \cite{LineVul2022}. From the interpretable machine-learning perspective, it
seems that the existing methods 
\cite{ribeiro2016should,shrikumar2017learning,lundberg2017unified,learning-to-explain-l2x} with adaptations can be applied. 
However, besides \textbf{L2X}  \cite{learning-to-explain-l2x},
none of the others is applicable to the context
of statement-level vulnerability detection. It is worth noting that unlike L2X, ICVH, and our proposed methods, LineVul does not work straightforwardly at the statement level. In particular, it works at the token level to find out the important weight (the attention score) of each code token, from the source code section input, contributing to the model's prediction. After obtaining the attention scores for all code tokens, the authors integrate those scores into statement scores to find out the vulnerable statements.

Similar to ICVH, we did not compare our method to \textbf{VulDeeLocator} \cite{zhenli2020VulDeeLocator}
because: i) it cannot work directly with the source code, which needs to be compiled to the Lower Level Virtual Machine code, and ii) it cannot be operated in the unsupervised setting because of requiring information relevant
to vulnerable code statements. About two other methods for statement-level vulnerability detection including \textbf{LineVD} \cite{LineVD-2022} and \textbf{VELVET} \cite{VELVET2022}, we also did not compare our method with these baselines because they cannot be operated in the unsupervised setting (i.e., this is a challenging setting and the main focus of our paper).
\subsection{Model Configuration}
\label{sec:model_configuration}
For the L2X \cite{learning-to-explain-l2x} and ICVH \cite{van-ijcnn2021}
methods, they were proposed to work as explaining models aiming to
explain the output of a learning model (i.e., which approximates the
true conditional distribution $p(Y\mid F)$). To use these methods
directly to deal with the problem of statement-level software vulnerability
detection, we keep their principles and apply them directly to approximate
$p(Y\mid F)$ using $p(Y\mid\tilde{F})$ where $\tilde{F}$ consists
of the selected vulnerability-relevant source code statements. To
these methods, for the architecture of the random selection network
obtaining $\tilde{F}$ as well as the classifier working on $\tilde{F}$
to mimic $p(Y\mid F)$, we follow the structures mentioned in the
corresponding original papers.

To our \ourapp~method, for the $\omega\left(\cdot;\alpha\right)$
and $g\left(\cdot;\beta\right)$ networks, we used deep feed-forward
neural networks having three and two hidden layers with the size of
each hidden layer in $\left\{ 100,300\right\} $. The dense hidden
layers are followed by a ReLU function as nonlinearity and Dropout
\cite{srivastava14a} with a retained fixed probability $p=0.8$
as regularization. The last dense layer of the $\omega\left(\cdot;\alpha\right)$
network for learning a discrete distribution is followed by a sigmoid
function while the last dense layer of the $g\left(\cdot;\beta\right)$ network is followed by a softmax function for predicting. The temperature $\nu$ for the Gumbel softmax distribution is equal
to $0.5$. The number of chosen clusters guiding the computation of the proposed clustered
spatial contrastive learning mentioned in Eq. (\ref{eq:cluster_spatial_cl})
is in {$\{{3},{5}, {7}\}$. The trade-off hyper-parameter $\alpha$ representing
the weight of the proposed clustered spatial contrastive learning
term in the final objective function (mentioned in Eq. (\ref{eq:final-objective-function}))
is in $\{10^{-2},10^{-1},10^{0}\}$ while the scalar temperature $\tau$
is equal to $0.5$. 
It is worth noting that the length ($L$) of each function is padded or truncated to $100$ code statements for the CWE-119 and CWE-399 datasets while to the Big-Vul dataset, $L$ is set to $150$. We chose these values based on the fact that for the CWE-119 and
CWE-399 datasets, over 90\% of the source code functions, have the
number of code statements less than or equal to 100 while this value
is equal to 150 for the Big-Vul dataset. Furthermore, almost all important information relevant to the vulnerability of each source
code function lies in the 100 and 150 first code statements for the
(CWE-119 and CWE-399) and Big-Vul datasets respectively.

For our \ourapp~method and baselines, we employed
the Adam optimizer \cite{KingmaB14} with an initial learning rate
in $\{10^{-3},10^{-4}\}$, while the mini-batch size is $100$. We split the data of each dataset into three random partitions.
The first partition contains 80\% for training, the second partition
contains 10\% for validation and the last partition contains 10\%
for testing. For each dataset, we used $10$ epochs for the training
process. We additionally applied gradient clipping regularization
to prevent over-fitting. For each method, we ran the corresponding
model $5$ times and reported the averaged VCP, VCA, Top10 ACC, and IFA\emph{ }measures. We ran our experiments in Python
using Tensorflow \cite{abadi2016tensorflow} for the used methods
on an Intel E5-2680, having 12 CPU Cores at 2.5 GHz with 128GB RAM, integrated NVIDIA Tesla K80.

For the LineVul method, we kept the proposed architecture (i.e., a
Transformer-based model) and used the shared source code written in
Python using Pytorch \cite{Pytorch2019} from the authors \cite{LineVul2022}.
\section{Experimental results}
\label{sec:exp_results}


\subsection*{RQ1: \rqone}
\smallsection{Approach}
We compared the performance of our \ourapp~method with baselines
including L2X \cite{learning-to-explain-l2x}, ICVH \cite{van-ijcnn2021},
and LineVul \cite{LineVul2022} in the unsupervised setting (i.e.,
we do not use any information about the ground truth of vulnerable code
statements in the training process) for localizing the vulnerable
code statements. We aim to find out the top $K$ statements that mostly
cause the vulnerability of each function. In this experiment,
the number of selected code statements for each function is fixed
equal to $10$ shown in Table \ref{tab:Performance-results-399-119-k10}.

\vspace{1mm}
\begin{table}[th]
\centering{}
\caption{Performance results in terms of the main measures for statement-level
vulnerability detection including VCP, VCA, Top-10 accuracy (Top-10
ACC), and IFA on the testing set of the CWE-399, CWE-119, and Big-Vul datasets for the L2X, ICVH, LineVul and \ourapp~
methods with $K=10$ (the best performance is shown in \textbf{bold}).\label{tab:Performance-results-399-119-k10}}
\vspace{-1mm}
\resizebox{0.85\columnwidth}{!}{
\begin{tabular}{|ccccccc|}
\hline 
Dataset & K & Method & VCP & VCA & Top-10 ACC & IFA\tabularnewline
\hline 
\multirow{4}{*}{CWE-399} & \multirow{4}{*}{10} & L2X \cite{learning-to-explain-l2x} & 88.5\% & 83.0\% & 83.0\% & 3.8\tabularnewline
 &  & ICVH \cite{van-ijcnn2021} & 84.5\% & 77.0\% & 81.0\% & 5.5\tabularnewline
 &  & LineVul \cite{LineVul2022} & 92.0\% & 89.0\% & 91.0\% & 3.8\tabularnewline
\cline{3-7} \cline{4-7} \cline{5-7} \cline{6-7} \cline{7-7} 
 &  & \ourapp~ (ours) & \textbf{96.6\%} & \textbf{95.0\%} & \textbf{95.0\%} & \textbf{2.4}\tabularnewline
\hline 
\multirow{4}{*}{CWE-119} & \multirow{4}{*}{10} & L2X \cite{learning-to-explain-l2x} & 93.2\% & 90.3\% & 94.1\% & 3.3\tabularnewline
 &  & ICVH \cite{van-ijcnn2021} & 93.5\% & 91.1\% & 94.5\% & 2.2\tabularnewline
 &  & LineVul \cite{LineVul2022} & 93.0\% & 89.0\% & 91.0\% & \textbf{2.0}\tabularnewline
\cline{3-7} \cline{4-7} \cline{5-7} \cline{6-7} \cline{7-7} 
 &  & \ourapp~ (ours) & \textbf{97.5\%} & \textbf{96.5\%} & \textbf{97.6\%} & 2.1\tabularnewline
\hline 
\multirow{4}{*}{Big-Vul} & \multirow{4}{*}{10} & L2X \cite{learning-to-explain-l2x} & 65.5\% & 60.9\% & 69.7\% & 2.2\tabularnewline
 &  & ICVH \cite{van-ijcnn2021} & 73.8\% & 69.6\% & 76.8\% & 2.2\tabularnewline
 &  & LineVul \cite{LineVul2022} & 62.0\% & 60.0\% & 74.0\% & 3.7\tabularnewline
\cline{3-7} \cline{4-7} \cline{5-7} \cline{6-7} \cline{7-7} 
 &  & \ourapp~ (ours) & \textbf{80.5\%} & \textbf{77.6\%} & \textbf{80.6\%} & \textbf{1.5}\tabularnewline
\hline 
\end{tabular}}
\vspace{-0mm}
\end{table}
\vspace{1mm}

\smallsection{Result}
The results in Table \ref{tab:Performance-results-399-119-k10}
show that our \textit{\ourapp~method achieved a much higher performance
on statement-level vulnerability detection measures}, including
VCP, VCA, and Top-10 ACC, compared to the L2X, ICVH, and LineVul methods
on the CWE-399, CWE-119, and Big-Vul datasets.
To the IFA measure, our \ourapp~method also achieved a higher performance
on the CWE-399 and Big-Vul datasets while for
the CWE-119 dataset, our \ourapp~method obtained a comparable value
to the highest one (i.e., 2.0) from the LineVul baseline.

Generally, \textit{our \ourapp~method obtained a higher performance on
the VCP, VCA, and Top-10 ACC measures} -- from 4\% to 14\% for the CWE-119
dataset, from 3\% to 7\% for the CWE-399 dataset, and 3\% to 11\%
for the Big-Vul dataset compared with the baselines.
For example, to the CWE-399 dataset with $K=10$, our \ourapp~method
achieved 96.6\% for VCP, 95.0\% for both VCA and Top-10 ACC, and 2.4
for IFA while (the L2X, ICVH, and LineVul methods) achieved (88.5\%, 84.5\%, and
92.0\%) for VCP, (83.0\%, 77.0\%, and 89.0\%) for VCA, (83.0\%, 81.0\%,
and 91.0\%) for Top-10 ACC, and (3.8, 5.5, and 3.8) for IFA respectively.

In addition to the main measures for statement-level vulnerability detection
including VCP, VCA, Top-10 accuracy, and IFA, we also computed the
function classification accuracy (ACC) for our \ourapp~method and
baselines (i.e., the L2X, ICVH, and LineVul methods) for the used datasets. With
$K=10$, our \ourapp~method and baselines all obtained an ACC higher than 96\%, 93\%, and 91\% for CWE-399, CWE-119, and Big-Vul datasets respectively.

\vspace{3mm}
\colorbox{gray!20}{
\begin{minipage}{0.88\textwidth}
\textbf{In conclusion for RQ1}: The experimental results (in Table \ref{tab:Performance-results-399-119-k10}) on the statement-level vulnerability detection measures (i.e., VCP, VCA, Top-10 accuracy (Top-10 ACC), and IFA) along with the function classification accuracy (ACC) show the superiority of our \ourapp~method in achieving a high performance in terms of making the label predictions and localizing the vulnerable code statements on the used real-word datasets over the baselines. 
\end{minipage}
}
\vspace{3mm}

\subsection*{RQ2: \rqtwo}
\smallsection{Approach}
By using the multivariate Bernoulli distribution in the selection
process, the ICVH and \ourapp~methods can be operated in\emph{
the semi-supervised setting} \cite{van-ijcnn2021} (i.e., where we assume that the core
vulnerable statements in a small proportion of functions are manually
annotated) in addition to the unsupervised setting. \revise{We can leverage such ground-truth
information by adding the maximization of a log-likelihood as an additional
training objective:}

{\small{}
\[
\max\,\{\sum_{k\in I_{c}}\log p_{k}+\sum_{k\notin I_{c}}\log(1-p_{k})\}
\]
}{\small\par}
\revise{
\noindent where $I_{c}=[i_{1},...,i_{m}]$. We then add the above
additional objective function to the final objective function (mentioned in Eq. (\ref{eq:final-objective-function})) with the trade-off parameter $\eta>0$ (i.e., we set the value of $\eta$ in $\{10^{-3},10^{-2},10^{-1}\}$).}

Here, we investigated the performance of our \ourapp~method in the semi-supervised setting compared with its performance in the unsupervised setting for highlighting vulnerable statements. We also compared the performance of our \ourapp~method with the
ICVH method in the semi-supervised setting. We conducted these experiments
on the CWE-119, CWE-399, and Big-Vul datasets. In the semi-supervised setting,
we assume that there is a small portion of the training set (i.e., 10\%) having the ground truth of vulnerable code statements.

\begin{table}[h]
\centering{}
\caption{Performance results of the ICVH and \ourapp~ methods with K=10
for the VCP, VCA\emph{,} TopK ACC, and IFA measures on the testing set of the CWE-399, CWE-119, and Big-Vul datasets in the unsupervised and semi-supervised
settings (with 10\% of the training set having the ground truth of vulnerable
code statements). In the semi-supervised setting, we denote our \ourapp~
method as \ourapp~-S10 while the ICVH method is denoted as ICVH-S10 (the best performance is shown in \textbf{bold}).
\label{tab:Performance-results-in-semi}}
\vspace{-1mm}
\resizebox{0.8\columnwidth}{!}{
\begin{tabular}{|ccccccc|}
\hline 
\multicolumn{1}{|c}{Dataset} & \multicolumn{1}{c}{K} & \multicolumn{1}{c}{Method} & \multicolumn{1}{c}{VCP} & \multicolumn{1}{c}{VCA} & \multicolumn{1}{c}{Top10 ACC} & IFA\tabularnewline
\hline 
\multirow{4}{*}{CWE-119} & \multirow{4}{*}{10} & ICVH \cite{van-ijcnn2021} & 93.5\% & 91.1\% & 94.5\% & 2.2\tabularnewline
 &  & ICVH-S10 \cite{van-ijcnn2021} & 99.4\% & 99.3\% & \textbf{100\%} & \textbf{1.2}\tabularnewline
\cline{3-7} \cline{4-7} \cline{5-7} \cline{6-7} \cline{7-7} 
 &  & \ourapp~ & 97.5\% & 96.5\% & 97.6\% & 2.1\tabularnewline
 &  & \ourapp~-S10 & \textbf{99.9\%} & \textbf{99.8\%} & \textbf{100\%} & 1.9\tabularnewline
\hline 
\multirow{4}{*}{CWE-399} & \multirow{4}{*}{10} & ICVH \cite{van-ijcnn2021} & 84.5\% & 77.0\% & 81.0\% & 5.5\tabularnewline
 &  & ICVH-S10 \cite{van-ijcnn2021} & 90.5\% & 86.00\% & 100\% & 5.0\tabularnewline
\cline{3-7} \cline{4-7} \cline{5-7} \cline{6-7} \cline{7-7} 
 &  & \ourapp~ & 96.6\% & 95.0\% & 95.0\% & 2.4\tabularnewline
 &  & \ourapp~-S10 & \textbf{99.3\%} & \textbf{99.0\%} & \textbf{99.0\%} & \textbf{2.0}\tabularnewline
\hline 
\multirow{4}{*}{Big-Vul} & \multirow{4}{*}{10} & ICVH \cite{van-ijcnn2021} & 73.8\% & 69.6\% & 76.8\% & 2.2\tabularnewline
 &  & ICVH-S10 \cite{van-ijcnn2021} & 79.3\% & 76.1\% & 80.6\% & 3.5\tabularnewline
\cline{3-7} \cline{4-7} \cline{5-7} \cline{6-7} \cline{7-7} 
 &  & \ourapp~ & 80.5\% & 77.6\% & 80.6\% & \textbf{1.5}\tabularnewline
 &  & \ourapp~-S10 & \textbf{82.9\%} & \textbf{80.6\%} & \textbf{83.6\%} & \textbf{1.5}\tabularnewline
\hline 
\end{tabular}} 
\vspace{-0mm}
\end{table}

\vspace{1mm}
\smallsection{Result}
The experimental results in Table \ref{tab:Performance-results-in-semi}
show that by using a small portion of data having the ground truth
(i.e., 10\%) of vulnerable code statements, the performance of our
\ourapp~method in the semi-supervised setting significantly increased
compared to its performance in the unsupervised setting for all the used datasets.

The results in Table \ref{tab:Performance-results-in-semi} also show
that the model's performance of the ICVH method in the semi-supervised
setting increased compared to its performance in the unsupervised
setting. However, our \ourapp~method still obtained a higher performance
on three used datasets (i.e., the CWE-119, CWE-399, and Big-Vul datasets) for most of the used metrics,
especially for the VCP and VCA measures.

\vspace{3mm}
\colorbox{gray!20}{
\begin{minipage}{0.88\textwidth}
\textbf{In conclusion for RQ2}: The experimental results (in Table \ref{tab:Performance-results-in-semi}) show a considerable increase in the model's performance of our \ourapp~ method in the semi-supervised setting compared to itself in the unsupervised setting. We also observed an improvement in the model's performance of the ICVH method. However, compared to ICVH, our proposed method still obtains a higher performance on
the used datasets for most metrics, especially for the VCP and VCA measures.
\end{minipage}
}
\vspace{3mm}
\subsection*{RQ3: \rqthree}
\smallsection{Approach}
We investigated the performance of our proposed method
in three different cases related to contrastive learning including i) using clustered spatial contrastive
learning mentioned in Eq. (\ref{eq:cluster_spatial_cl}) (\ourapp~-with-CSCL), ii) using normal contrastive learning mentioned
in Eq. (\ref{eq:cl_spatial_cl}) (\ourapp~-with-CL,
and iii) not using contrastive learning (\ourapp~-without-CL) (i.e., in this case, we only use the mutual information for guiding the whole training process).

\vspace{1mm}
\smallsection{Result}
The experimental results in Table
\ref{tab:cluster_cl_vs_cl-1} show that our method using
clustered spatial contrastive learning
obtains a much higher performance compared to its performance when
using normal contrastive learning, especially compared to the case not using contrastive learning on the VCP, VCA, Top-10 ACC, and
IFA measures. These results
demonstrate the efficiency and superiority of our proposed clustered
spatial contrastive learning in modeling the important properties
for the relationship of vulnerable patterns between the source code
sections in order to improve the selection process of $\tilde{F}$.

\begin{table}[H]
\centering{}
\caption{Performance results on VCP, VCA, Top-10 ACC, and IFA measures for the testing set of the CWE-399, CWE-119, and Big-Vul datasets
for our method in three different cases including
\ourapp~-with-CSCL, \ourapp~-with-CL, and \ourapp~-without-CL with
$K=10$ (the best performance is shown in \textbf{bold}).\label{tab:cluster_cl_vs_cl-1}}
\vspace{0mm}
\resizebox{0.7\columnwidth}{!}{
\begin{tabular}{|ccccccc|}
\hline 
Dataset & K & Method & VCP & VCA & Top-10 ACC & IFA\tabularnewline
\hline 
\multirow{3}{*}{CWE-399} & \multirow{3}{*}{10} & \ourapp~-without-CL & 85.8\% & 79.0\% & 80.0\% & 4.5\tabularnewline
 &  & \ourapp~-with-CL & 91.2\% & 87.0\% & 87.0\% & 3.5\tabularnewline
\cline{3-7} \cline{4-7} \cline{5-7} \cline{6-7} \cline{7-7} 
 &  & \ourapp~-with-CSCL & \textbf{96.6\%} & \textbf{95.0\%} & \textbf{95.0\%} & \textbf{2.4}\tabularnewline
\hline 
\multirow{3}{*}{CWE-119} & \multirow{3}{*}{10} & \ourapp~-without-CL & 92.1\% & 89.2\% & 93.6\% & 4.1\tabularnewline
 &  & \ourapp~-with-CL & 94.5\% & 92.3\% & 94.2\% & 2.6\tabularnewline
\cline{3-7} \cline{4-7} \cline{5-7} \cline{6-7} \cline{7-7} 
 &  & \ourapp~-with-CSCL & \textbf{97.5\%} & \textbf{96.5\%} & \textbf{97.6\%} & \textbf{2.1}\tabularnewline
\hline
\multirow{3}{*}{Big-Vul} & \multirow{3}{*}{10} & \ourapp~-without-CL & 74.3\% & 70.1\% & 76.1\% & 2.3\tabularnewline
 &  & \ourapp~-with-CL & 75.9\% & 70.6\% & 76.5\% & 2.1\tabularnewline
\cline{3-7} \cline{4-7} \cline{5-7} \cline{6-7} \cline{7-7} 
 &  & \ourapp~-with-CSCL & \textbf{80.5\%} & \textbf{77.6\%} & \textbf{80.6\%} & \textbf{1.5}\tabularnewline
\hline 
\end{tabular}}
\vspace{0mm}
\end{table}

\vspace{2mm}
\colorbox{gray!20}{
\begin{minipage}{0.93\textwidth}
\textbf{In conclusion for RQ3}: The experimental results (in Table \ref{tab:cluster_cl_vs_cl-1}) prove the effectiveness of our proposed cluster spatial contrastive learning described in Eq.(\ref{eq:cluster_spatial_cl}) in boosting the data representation learning and in figuring out the hidden
vulnerability pattern. In particular, our method using clustered spatial contrastive learning obtains a much higher performance compared to its performance when using normal contrastive learning, especially compared to the case of not using contrastive learning on the VCP, VCA, Top-10 ACC, and IFA measures.
\end{minipage}
}

\vspace{2mm}
\section{Discussion}

\subsection{Effect of the number of clusters} We investigated the correlation between the number
of chosen clusters guiding the computation of the proposed clustered
spatial contrastive learning term mentioned in Eq. (\ref{eq:cluster_spatial_cl}) and the
VCP and VCA measures for our \ourapp~method on the testing set of CWE-119,
CWE-399, and Big-Vul datasets. 

As shown in Figure \ref{fig:The-correlation-clusters-trade-off-1}, we observe that our \ourapp~method obtains a
higher performance for the VCP and VCA measures when the chosen cluster is in $\{3,5,7\}$, compared
to the case in which the chosen cluster is equal to $1$ or $10$.
In the case of the chosen cluster equal to $1$, we assume that to
each vulnerability type (e.g., the buffer overflow error), there is
only one dynamic pattern causing the corresponding vulnerability;
however, in reality, for each vulnerability type, there are some vulnerability
patterns.
In the case when we set the chosen cluster equal to $10$, we may
set the number of vulnerability patterns higher than the true one.
These are the reasons why the model's performance in these cases is lower than the case when the chosen cluster varies in $\{3,5,7\}$ which can reflect more appropriate values of the true number of vulnerability patterns.

\begin{figure}[H]%
\begin{centering}
\vspace{-5mm}
\includegraphics[width=0.65\columnwidth]{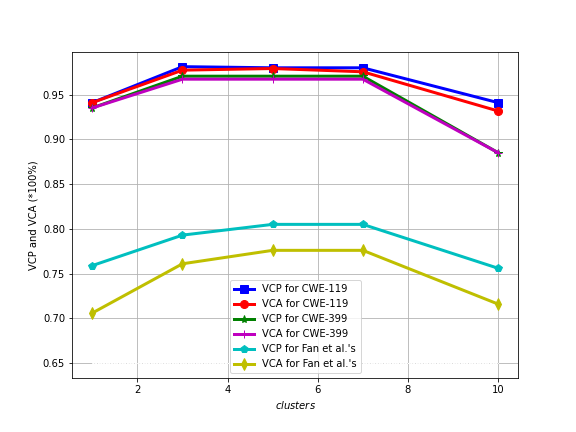}
\par\end{centering}\vspace{-5mm}
\caption{The correlation between the number of chosen clusters and the VCP
and VCA measures.\label{fig:The-correlation-clusters-trade-off-1}}
\end{figure}%
\vspace{-5mm}

\subsection{Parameter sensitivity} 
We investigated the correlation between
the hyper-parameter $\alpha$ which represents the weight
of the proposed clustered spatial contrastive learning term in the
final objective function (\ref{eq:final-objective-function}) and the
VCP and VCA measures for our \ourapp~ method on the testing set of the CWE-399 and CWE119 datasets in the unsupervised setting.

The results
in Figure \ref{fig:The-correlation-clusters-trade-off} show that we can obtain a better model's performance when
$\alpha$ varies in $\{10^{0},2\times10^{0},3\times10^{0}\}$, compared
to the case in which $\alpha$ varies in $\{10^{-3},10^{-2},10^{-1}\}$.
That indicates the importance of the clustered spatial contrastive learning term in the training process to model important properties for the relationship of vulnerable patterns between the source code sections. That helps boost the selection process of $\tilde{F}$ to improve the model's performance significantly. Furthermore, the results in Figure \ref{fig:The-correlation-clusters-trade-off} also indicate that we should set the value
of the trade-off hyper-parameter $\alpha$ higher than $10^{-1}$
to make sure that we use enough information of the clustered spatial
contrastive learning term to enhance the representation learning and
robust the selection process of vulnerability-relevant code statements.

\begin{figure}[h]%
\begin{centering}
\vspace{-6mm}
\includegraphics[width=0.65\columnwidth]{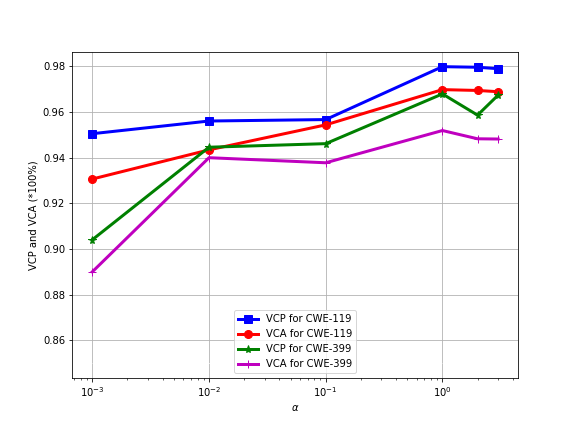}\vspace{-5mm}
\par\end{centering}
\caption{The correlation between the trade-off
hyper-parameter $\alpha$ and the VCP and VCA measures on the testing set of the CWE-399 and CWE119 datasets.\label{fig:The-correlation-clusters-trade-off}}
\vspace{-3mm}
\end{figure}%

\subsection{Explanatory capability of our proposed method}

In order to demonstrate the ability of our proposed method in detecting
the vulnerable code patterns and statements in the vulnerable functions to support
security analysts and code developers, in this section, we show a
visualization of the selected code statements for a vulnerable function in the unsupervised setting.


As shown in Figure \ref{fig:True-labels-and-5-2}, the two functions in the left-hand figure (a) share some similar flaws. These include (i) a potential flaw of reading data from the console, i.e., in the statement \emph{"if ( fgets ( var2 , var3 , stdin ) != NULL )"} of the first function and in the statement \emph{fscanf ( var3 , str , \& var1 ) ;} of the second function, and (ii) a potential flaw due to no maximum limitation for memory allocation in the statements \emph{"if ( var1 > wcslen ( var9 ) )"} and \emph{"if ( var1 > wcslen ( var5 ) )"} of the first and second functions, respectively. The two functions on the right-hand figure (b) also share some similar vulnerabilities including (i) a potential flaw of reading data from an environment, i.e., in the statement \emph{"wcsncat ( var1 + var2 , var3 , 100 - var2 - 1 ) ;"} of the first function and in the statement \emph{"if ( fgetws ( var1 + var2 , ( int ) ( 100 - var2 ) , var3 ) != var4 )"} of the second function, and (ii) a flaw of incrementing the pointer in the loop which will cause us to free the memory block not at the start of the buffer in the statement \emph{"for ( ; * var1 != str ; var1 ++ )"} of both these functions.

\begin{figure*}[ht]%
\begin{centering}
\vspace{1mm}
\begin{tabular}{c}
\includegraphics[width=0.98\textwidth]{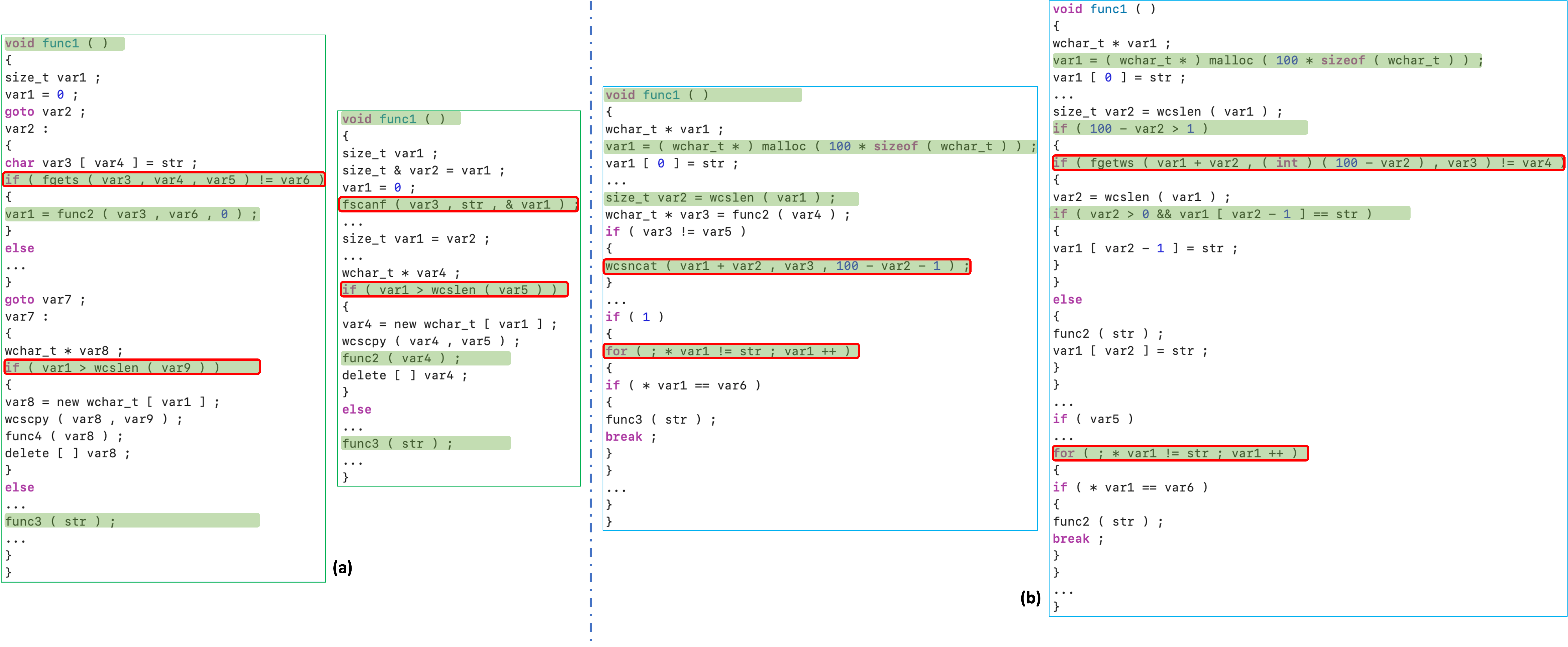}\tabularnewline
\end{tabular}\vspace{-3mm}
\par\end{centering}
\caption{An example showing the real-world source code functions and the vulnerability-relevant selected code statements with $K=5$ (i.e., the green background lines) from our \ourapp~ method. Note that the selected statements in each function with the red border specify the core vulnerable statements
obtained from the ground truth and also detected by our method. Two functions on the left-hand figure (a) consist of two similar potential flaws which are \textit{"reading data from the console"} and \textit{"no maximum limitation for memory allocation"}. Two functions on the right-hand figure (b) share a similar potential flaw and flaw including \textit{"reading data from an environment"} and \textit{"incrementing the pointer in the loop"}, respectively. For the demonstration purpose and simplicity, we choose simple and short vulnerable source code functions in which some statements are omitted and replaced by "..." for brevity. \label{fig:True-labels-and-5-2}}
\vspace{-3mm}
\end{figure*}%

In these functions, the selected
code statements from the model relevant to vulnerabilities are shown with $K=5$. The green background lines in each function highlight the detected code
statements while the green background lines with the red border specify the core vulnerable statements
obtained from the ground truth, and these lines are also detected by our method. \textbf{Our \ourapp~ method with $K=5$ can detect all of these potentially vulnerable code statements that make the corresponding functions vulnerable. It is worth noting that to the source code sections having similar vulnerability patterns, the subset of vulnerability-relevant selected statements (i.e., $F^{top}$) from our \ourapp~method are quite similar. That further demonstrates the ability of our method in detecting vulnerable code patterns and statements in the vulnerable functions}.

\subsection{Additional experiments}
\paragraph{\textbf{Auxiliary measure}}

As mentioned in the experiments section of the paper, the main purpose
of our \ourapp~ method is to support security analysts and code
developers to narrow down the vulnerable scope for seeking vulnerable
statements. This would be helpful in the context that they need to
identify several vulnerable statements from hundreds or thousands
of lines of code.

We aim to specify lines of statements (e.g., \emph{top-K=10}) so that with a high probability those lines cover most or all
vulnerable statements. Bearing this incentive, and inspired by \cite{van-ijcnn2021},
to evaluate the performance of our proposed method and baselines,
we use two main measures introduced in \cite{van-ijcnn2021} including: \emph{vulnerability coverage proportion (VCP)} (i.e., the proportion
of correctly detected vulnerable statements over all vulnerable statements
in a dataset) and \emph{vulnerability coverage accuracy (VCA)} (i.e.,
the ratio of the successfully detected functions, having all vulnerable
statements successfully detected, over all functions in a dataset).
We also apply using two other measures including \emph{Top-10 Accuracy}
(i.e., it measures the percentage of vulnerable functions where at
least one actual vulnerable lines appear in the top-10 ranking) and
Initial False Alarm (\emph{IFA}) (i.e., it measures the number of
incorrectly predicted lines (i.e., non-vulnerable lines incorrectly
predicted as vulnerable or false alarms) that security analysts need
to inspect until finding the first actual vulnerable line for a given
function) used in \cite{LineVul2022}.

In practice, security analysts and code developers can set their preferable
\emph{top-K} for the used methods, so we need a measure that
penalizes large \emph{top-K}. To this end, we propose an auxiliary
measure named VCE (i.e., vulnerable coverage efficiency) which measures
the percentage of vulnerable statements detected over the number of
selected statements. For example, if a source code section has 3 core
vulnerable code statements and using \emph{top-K}=5, we can successfully
detect 2 vulnerable statements. The VCE measure in this case is equal
to 2/5 = 0.4. This additional measure would offer a helpful measure
of efficiency to users.

We computed the auxiliary VCE measure for our \ourapp~ method
and baselines (i.e., L2X, ICVH, and LineVul) with \emph{top-K}=5 in the
unsupervised setting. The experimental results mentioned in Table
\ref{tab:Experimental-results-in-VCE} show that our \ourapp~ method
obtained the highest VCE measure in both CWE-399 and CWE-119 datasets
compared to the baselines. In particular, to the CWE-399 dataset,
our \ourapp~ method gained 45.1\% for the auxiliary VCE measure.
It means that in this case, we can detect 5$\times$0.451 = 2.255
vulnerable statements out of 5 spotted lines.

\begin{table}[h]%
\vspace{1mm}
\centering{}
\caption{Experimental results in terms of the auxiliary VCE measure on the testing set of the CWE-399 and CWE-119 datasets for the L2X, ICVH,
LineVul, and \ourapp~ methods with $K=5$ in the unsupervised setting (the best performance is shown in \textbf{bold}).\label{tab:Experimental-results-in-VCE}}
\resizebox{0.65\columnwidth}{!}{
\begin{tabular}{|cccc|}
\hline 
Dataset & K & Method & VCE\tabularnewline
\hline 
\multirow{4}{*}{CWE-399} & \multirow{4}{*}{5} & L2X \cite{learning-to-explain-l2x} & 43.6\%\tabularnewline
 &  & ICVH \cite{van-ijcnn2021} & 39.1\%\tabularnewline
 &  & LineVul \cite{LineVul2022} & 29.2\%\tabularnewline
\cline{3-4} \cline{4-4} 
 &  & \ourapp~ (ours) & \textbf{45.1\%}\tabularnewline
\hline 
\multirow{4}{*}{CWE-119} & \multirow{4}{*}{5} & L2X \cite{learning-to-explain-l2x} & 30.5\%\tabularnewline
 &  & ICVH \cite{van-ijcnn2021} & 26.4\%\tabularnewline
 &  & LineVul \cite{LineVul2022} & 23.9\%\tabularnewline
\cline{3-4} \cline{4-4} 
 &  & \ourapp~ (ours) & \textbf{31.1\%}\tabularnewline
\hline 
\end{tabular}}
\end{table}%

\subsection{Threats to Validity}
\label{sec:threats}
\paragraph{\textbf{Construct Validity}}
Key construct validity threats are if our assessments of the methods demonstrate the ability for detecting vulnerable code patterns and statements in the vulnerable functions in unsupervised and semi-supervised settings. The main purpose of our \ourapp~ method is to support security analysts and code developers to narrow down the vulnerable scope for seeking vulnerable statements. This would be helpful in the context that they need to identify several vulnerable statements from hundreds or thousands of lines of code in functions or programs. To evaluate the performance of our \ourapp~method and baselines, we use four main measures including: \emph{vulnerability coverage proportion (VCP)}, \emph{vulnerability coverage accuracy (VCA)}, \emph{Top-10 Accuracy}, and Initial False Alarm (\emph{IFA}) as mentioned in Section \ref{subsec:Measures-and-Evaluation}.

\vspace{-1mm}
\paragraph{\textbf{Internal validity}}
Key internal validity threats are relevant to the choice of hyper-parameter settings (i.e., optimizer, learning rate, number of layers in deep neural networks, etc.) as described in Section \ref{sec:model_configuration}. It is worth noting that finding a set of optimal hyperparameter settings of deep neural networks is expensive due to a large number of trainable parameters. To train our method, we only use the common and default values of hyper-parameters such as using Adam optimizer and the learning rate in $\{10^{-3},10^{-4}\}$. We also report the hyperparameter settings in the released reproducible source code to support future replication studies.

\vspace{-1mm}
\paragraph{\textbf{External validity}}
Key external validity threats include whether our proposed method will generalize to other vulnerabilities and whether they will work on other source code datasets. We mitigated this problem by using big and common but different vulnerabilities on three real-world source code datasets, namely, the CWE-399, CWE-199, and Big-Vul datasets.

\section{Related Work}
\label{sec:related_work}
Deep learning has been applied successfully to source code and binary
software vulnerability detection (SVD) 
\cite{Dam2017,jun_2018,VulDeePecker2018,le2019maximal,Zhuang2020,vancan2020,vancpn2020,Zhen2021, ReGVD2021, Boosting2021,nguyen2022cross,fu2022vulrepair,fu2023vulexplainer,fu2024vision}.
However, most of the current approaches only detect vulnerabilities at
either the function or program level, not at the more fine-grained code statement level. 
From the interpretable machine-learning perspective, it seems that
the existing methods \cite{ribeiro2016should,shrikumar2017learning,lundberg2017unified,learning-to-explain-l2x}
with adaptations can be ready to apply. 
However, besides L2X  \cite{learning-to-explain-l2x},
none of the others is applicable to the context
of statement-level vulnerability detection.

Recently, there have been several approaches \cite{zhenli2020VulDeeLocator,van-ijcnn2021,IVDetect2021,LineVul2022} proposed to solve the statement-level SVD problem. In particular,  \cite{zhenli2020VulDeeLocator} proposed VulDeeLocator, a deep learning-based method, requiring compiling the source code to Lower Level Virtual Machine code (this method cannot be used if a function cannot be compiled) and the information relevant to vulnerable code statements (the method cannot work in the unsupervised setting). Hoppity \cite{Dinella2020Hoppity} is a learning-based approach that uses a graph neural network (GNN) \cite{Scarselli2009ku} to detect (and fix bugs) at the token level in Javascript programs. However, the bug detection part in Hoppity needs to use the information relevant to the vulnerable code tokens and cannot work in the unsupervised setting.

\cite{van-ijcnn2021} proposed ICVH based on mutual information and used it as an explaining model to explain the reference model (i.e., the learning model approximating the true conditional distribution $p(Y\mid F)$). \cite{IVDetect2021} introduced the IVDetect method using a Feature-attention
Graph Convolution Network approach to predict function level vulnerabilities
and a GNNExplainer to identify which sub-graph contributed the most
to the predictions to locate the fine-grained location of vulnerabilities. However, such sub-graphs still contain many lines of code.
\cite{LineVul2022} proposed LineVul based on BERT \cite{Bert2018} to locate
vulnerable lines via attention weights. Recently, \cite{LineVD-2022} propose a deep learning framework, named LineVD, using the graph attention network (GAT) model \cite{velickovic2018graph}, formulating statement-level vulnerability detection as a node classification task for the statement level vulnerability detection. \cite{VELVET2022} introduce an ensemble learning approach, named VELVET, to locate vulnerable statements. In particular, the model combines graph-based and sequence-based neural networks to capture the context of a program graph and understand code semantics and vulnerable patterns. While ICVH \cite{van-ijcnn2021} and LineVul \cite{LineVul2022} can be operated in the unsupervised setting, LineVD and VELVET need to use the ground truth of vulnerable code statements during the training process via the supervised setting.

\vspace{-1mm}
\section{Conclusion}
\vspace{1mm}
\label{sec:conclusion}
We have proposed a novel end-to-end deep
learning-based approach for tackling the statement-level source code vulnerability
detection problem. In particular, we first leverage mutual information in learning a set of independent Bernoulli latent variables
that can represent the relevance of the source code statements to
the corresponding function's vulnerability.
We then proposed novel clustered
spatial contrastive learning in order to further improve the representation
learning and the robust selection process $\varepsilon$. Specifically,
our novel clustered spatial contrastive learning guides the
selection process $\varepsilon$ to select the hidden vulnerability
pattern characterized by $F^{top}$ in source code sections so that
the vulnerable source code sections originated from the same vulnerability
pattern are encouraged to have similarly selected vulnerability-relevant
code statements. Our experimental results on three real-world datasets show the superiority of our \ourapp~ method
over other state-of-the-art baselines in detecting
the vulnerable code statements in source code functions in both the unsupervised
and semi-supervised settings.

\bibliographystyle{ACM-Reference-Format}
\bibliography{reference}

\end{document}


\title{Statement-Level Vulnerability Detection: Learning Vulnerability Patterns Through Information Theory and Contrastive Learning
(Supplementary Material)
}

\renewcommand{\shorttitle}{Statement-Level Vulnerability Detection Through Information Theory and Contrastive Learning}

\author{Van Nguyen}
\affiliation{%
  \institution{Monash University}
  \city{Clayton}
  \country{Australia}}
\email{van.nguyen1@monash.edu}

\author{Trung Le}
\affiliation{%
  \institution{Monash University}
  \city{Clayton}
  \country{Australia}}
\email{trunglm@monash.edu}

\author{Chakkrit Tantithamthavorn}
\affiliation{%
  \institution{Monash University}
  \city{Clayton}
  \country{Australia}}
\email{chakkrit@monash.edu}

\author{Michael Fu}
\affiliation{%
  \institution{Monash University}
  \city{Clayton}
  \country{Australia}}
\email{yeh.fu@monash.edu}

\author{John Grundy}
\affiliation{%
  \institution{Monash University}
  \city{Clayton}
  \country{Australia}}
\email{john.grundy@monash.edu}


\author{Paul Quirk}
\affiliation{%
  \institution{Defence Science and Technology Group}
  \city{Adelaide}
  \country{Australia}}
\email{paul.quirk@defence.gov.au}

\author{Dinh Phung}
\affiliation{%
  \institution{Monash University}
  \city{Clayton}
  \country{Australia}}
\email{dinh.phung@monash.edu}

\renewcommand{\shortauthors}{Van Nguyen et al.}

\maketitle

\section{Additional experiments}
\subsection{Auxiliary measure}

As mentioned in the experiments section of the paper, the main purpose
of our \ourapp~ method is to support security analysts and code
developers to narrow down the vulnerable scope for seeking vulnerable
statements. This would be helpful in the context that they need to
identify several vulnerable statements from hundreds or thousands
of lines of code.

We aim to specify lines of statements (e.g., \emph{top-K=10}) so that with a high probability those lines cover most or all
vulnerable statements. Bearing this incentive, and inspired by \cite{van-ijcnn2021},
to evaluate the performance of our proposed method and baselines,
we use two main measures introduced in \cite{van-ijcnn2021} including: \emph{vulnerability coverage proportion (VCP)} (i.e., the proportion
of correctly detected vulnerable statements over all vulnerable statements
in a dataset) and \emph{vulnerability coverage accuracy (VCA)} (i.e.,
the ratio of the successfully detected functions, having all vulnerable
statements successfully detected, over all functions in a dataset).
We also apply using two other measures including \emph{Top-10 Accuracy}
(i.e., it measures the percentage of vulnerable functions where at
least one actual vulnerable lines appear in the top-10 ranking) and
Initial False Alarm (\emph{IFA}) (i.e., it measures the number of
incorrectly predicted lines (i.e., non-vulnerable lines incorrectly
predicted as vulnerable or false alarms) that security analysts need
to inspect until finding the first actual vulnerable line for a given
function) used in \cite{LineVul2022}.

In practice, security analysts and code developers can set their preferable
\emph{top-K} for the used methods, so we need a measure that
penalizes large \emph{top-K}. To this end, we propose an auxiliary
measure named VCE (i.e., vulnerable coverage efficiency) which measures
the percentage of vulnerable statements detected over the number of
selected statements. For example, if a source code section has 3 core
vulnerable code statements and using \emph{top-K}=5, we can successfully
detect 2 vulnerable statements. The VCE measure in this case is equal
to 2/5 = 0.4. This additional measure would offer a helpful measure
of efficiency to users.

We computed the auxiliary VCE measure for our \ourapp~ method
and baselines (i.e., L2X, ICVH, and LineVul) with \emph{top-K}=5 in the
unsupervised setting. The experimental results mentioned in Table
\ref{tab:Experimental-results-in-VCE} show that our \ourapp~ method
obtained the highest VCE measure in both CWE-399 and CWE-119 datasets
compared to the baselines. In particular, to the CWE-399 dataset,
our \ourapp~ method gained 45.1\% for the auxiliary VCE measure.
It means that in this case, we can detect 5$\times$0.451 = 2.255
vulnerable statements out of 5 spotted lines.

\begin{table}[h]%
\vspace{1mm}
\centering{}
\resizebox{0.65\columnwidth}{!}{
\begin{tabular}{|cccc|}
\hline 
Dataset & K & Method & VCE\tabularnewline
\hline 
\multirow{4}{*}{CWE-399} & \multirow{4}{*}{5} & L2X \cite{learning-to-explain-l2x} & 43.6\%\tabularnewline
 &  & ICVH \cite{van-ijcnn2021} & 39.1\%\tabularnewline
 &  & LineVul \cite{LineVul2022} & 29.2\%\tabularnewline
\cline{3-4} \cline{4-4} 
 &  & \ourapp~ (ours) & \textbf{45.1\%}\tabularnewline
\hline 
\multirow{4}{*}{CWE-119} & \multirow{4}{*}{5} & L2X \cite{learning-to-explain-l2x} & 30.5\%\tabularnewline
 &  & ICVH \cite{van-ijcnn2021} & 26.4\%\tabularnewline
 &  & LineVul \cite{LineVul2022} & 23.9\%\tabularnewline
\cline{3-4} \cline{4-4} 
 &  & \ourapp~ (ours) & \textbf{31.1\%}\tabularnewline
\hline 
\end{tabular}}
\vspace{1mm}
\caption{Experimental results in terms of the auxiliary VCE measure on the
testing set of the CWE-399 and CWE-119 datasets for the L2X, ICVH,
LineVul, and \ourapp~ methods with $K=5$ in the unsupervised setting (the best performance is shown in \textbf{bold}).\label{tab:Experimental-results-in-VCE}}
\end{table}%

\section{Released source code}
In this section, we summarize the information of datasets used in
our experiments, the required libraries (packages), and the instructions for reproducing
the experimental results of our proposed \ourapp~ method.

\subsection{Setup}

\paragraph{Datasets}

We used three real-world datasets including the CWE-399 dataset with
1,010 and 1,313 vulnerable/non-vulnerable functions for resource management
error vulnerabilities, the CWE-119 dataset with 5,582 and 5,099 vulnerable/non-vulnerable
functions for the buffer error vulnerabilities, and a big C/C++ dataset, namely Big-Vul,
provided by \cite{Bigdata2020} containing many types of vulnerabilities
such as Out-of-bounds Write, Improper Input Validation, and Path Traversal. 

For the CWE-399 and CWE-199 datasets collected by \cite{VulDeePecker2018},
we used the ones processed by \cite{van-ijcnn2021}. Additionally,
the Big-Vul dataset, considered as one of the
largest vulnerability datasets, includes the ground truth at the
statement level. The dataset is collected from 348 open-source Github
projects from 2002 to 2019. It consists of 188,636 C/C++ source code
functions where a ratio of vulnerable functions is 5.7\% (i.e., 10,900
vulnerable functions).

\paragraph{Requirements}

We implement our \ourapp~ method using Tensorflow \cite{abadi2016tensorflow}
(version 2.5), Python (version 3.8). Other required packages are scikit-learn,
numpy, scipy and pickle.

\subsection{Running source code samples}

Here, we provide the instructions for using the source code samples
of our \ourapp~ method on the Big-Vul dataset. Please download
the source code samples and the dataset via \newline \href{https://drive.google.com/drive/folders/1nmKrWB71fO_by0NFzjfRC2D9Jk7oE2FD?usp=sharing}{https://drive.google.com/drive/folders/1nm}.

\paragraph{\textbf{Folders and files}}
\begin{itemize}
\item The folder named "BigVul\_dataset" consists of all of the
files containing the Big-Vul dataset.
\vspace{1mm}
\item The file named "BigVul\_train\_evaluate.py" is the source code
for our proposed \ourapp~ method in both training and evaluating
processes for the Big-Vul dataset.
\vspace{1mm}
\item The file named "BigVul\_VCP\_VCA\_TopK\_IFA.py" is the source code
for computing the  VCP, VCA, Top-10 accuracy (Top-10 ACC), and IFA measures for the Big-Vul dataset.
\vspace{1mm}
\item The file named "Utils.py" is a collection of supported Python functions used in the training and evaluating
processes of the model.
\end{itemize}

\paragraph{\textbf{Train, evaluate and get the results}}
\begin{itemize}
\item To train our model, please use the following command, for example, \newline
"\emph{python BigVul\_train\_evaluate.py -{}-lr=1e-4 -{}-sigma=1e-1
-{}-tau=0.5 -{}-temp=0.5 -{}-dim\_dnn=300 -{}-clusters=7 -{}-train\_epochs=10
-{}-do\_train -{}-home\_dir=./BigVul\_results/}"
\vspace{1mm}
\item To evaluate our \ourapp~ method, please use
the following command, for example: \newline "\emph{python BigVul\_train\_evaluate.py
-{}-lr=1e-4 -{}-sigma=1e-1 -{}-tau=0.5 -{}-temp=0.5 -{}-dim\_dnn=300
-{}-clusters=7 -{}-do\_eval -{}-home\_dir=./BigVul\_results/}"
\vspace{1mm}
\item To get the results of the main measures including VCP, VCA, Top-10 accuracy (Top-10 ACC), and IFA,
please use the file named "BigVul\_VCP\_VCA\_TopK\_IFA.py",
with the following command, for example, \newline "\emph{python BigVul\_VCP\_VCA\_TopK\_IFA.py -{}-home\_dir=./BigVul\_results/}".
\end{itemize}

\paragraph{\textbf{The model configuration}}

For the \ourapp~ model configuration, please read the "Model configuration" section in the main paper.


\bibliographystyle{ACM-Reference-Format}
\bibliography{reference}